\documentclass[aps,preprint,epsf,superscriptaddress,nofootinbib]{revtex4-1}
\pdfoutput=1
\usepackage{graphicx,epsfig,amssymb,amsmath,color,hyperref,cancel}
\newcommand{\os}{{\rm OS}}
\newcommand{\msbar}{{\rm \overline{MS}}}
\allowdisplaybreaks

\begin{document}
\draft \tighten
\title{High Scale Validity of the DFSZ Axion Model with Precision}

\author{Satsuki Oda}
\email{satsuki.oda@oist.jp}
\affiliation{Okinawa Institute of Science and Technology Graduate
University (OIST), Onna, Okinawa 904-0495, Japan} \affiliation{Institute for Pacific Rim Studies, Meio University, Nago, Okinawa 905-8585, Japan}
\author{Yutaro Shoji}
\email{yshoji@kmi.nagoya-u.ac.jp}
\affiliation{Kobayashi-Maskawa Institute for the Origin of Particles and
the Universe, Nagoya University, Nagoya, Aichi 464-8602, Japan}
\author{Dai-suke Takahashi}
\email{daisuke.takahashi@oist.jp}
\affiliation{Okinawa Institute of Science and Technology Graduate
University (OIST), Onna, Okinawa 904-0495, Japan} \affiliation{Institute for Pacific Rim Studies, Meio University, Nago, Okinawa 905-8585, Japan}

\begin{abstract}
With the assumption of classical scale invariance at the Planck scale,
the DFSZ axion model can generate the Higgs mass terms of the
appropriate size through technically natural parameters and may be valid
up to the Planck scale.  We discuss the high scale validity of the Higgs
sector, namely the absence of Landau poles and the vacuum stability. The
Higgs sector is identical to that of the type-II two Higgs doublet model
with a limited number of the Higgs quartic couplings. We utilize the
state-of-the-art method to calculate vacuum decay rates and find that
they are enhanced at most by $10^{10}$ compared with the tree level
evaluation. We also discuss the constraints from flavor observables,
perturbative unitarity, oblique parameters and collider searches. We
find that the high scale validity tightly constrains the parameter
region, but there is still a chance to observe at most about $10\%$
deviation of the $125~{\rm GeV}$ Higgs couplings to the fermions.
\end{abstract}
\maketitle

\section{Introduction}
An invisible axion
\cite{Peccei:1977hh,Weinberg:1977ma,Wilczek:1977pj,Abbott:1982af,Dine:1982ah,Preskill:1982cy,Kuster:2008zz,Shifman:1979if,Kim:1979if,Zhitnitsky:1980tq,Dine:1981rt}
is one of the plausible solutions to the strong CP problem and is also
an excellent dark matter candidate. We focus on the DFSZ axion model
\cite{Zhitnitsky:1980tq,Dine:1981rt}, where the standard model (SM) is
extended with a SM singlet complex scalar and an additional Higgs
doublet.  Since the Higgs doublets have non-zero Peccei-Quinn (PQ)
charges, the Higgs couplings are tightly restricted by the PQ
symmetry. For example, dangerous flavor changing neutral currents (FCNC)
are forbidden at the tree level and the CP is not broken spontaneously
in the scalar sector.

In this paper, we discuss the possibility that the DFSZ axion model
remains valid up to the Planck scale. In such a scenario, one of the
disadvantages is that we need to give up a complete explanation of the
hierarchy between the Planck scale and the electroweak (EW)
scale. However, if there is a mechanism that realizes classical scale
invariance at the Planck scale, the hierarchy problem may be solved
without introducing supersymmetry or compositeness
\cite{Bardeen:1995kv,Hill:2005wg,Aoki:2012xs}. Since the scale
invariance is violated at the quantum level, the PQ breaking scale can
appear through the dimensional transmutation. If the PQ sector and the
Higgs sector are connected by (technically natural) tiny couplings, the
PQ breaking can also generate the Higgs mass terms without causing a
hierarchy problem \cite{Allison:2014hna,Allison:2014zya}. Since the PQ
breaking sector decouples from the Higgs sector due to the tiny
couplings, the model is well approximated by the type-II two Higgs
doublet model (THDM) with a restricted number of coupling
constants. Importantly, the additional Higgs bosons should be around the
EW scale in this scenario since there is no technically natural
parameter that accommodates a hierarchy among the Higgs boson masses.

Another disadvantage is that the model does not explain the neutrino
masses, the baryon asymmetry of the Universe, or inflation. However,
they can be explained without affecting the Higgs sector. For example,
one may consider the see-saw mechanism with right handed neutrinos
having a few orders of magnitude smaller masses than the PQ breaking
scale \cite{Vissani:1997ys,Farina:2013mla,Foot:2013hna}. It can explain
the neutrino masses and also the baryon asymmetry of the Universe with
$\tan\beta\gtrsim4$ \cite{Clarke:2015hta}. However, it does not cause
the hierarchy problem thanks to the tiny Yukawa couplings of the right
handed neutrinos. As for inflation, one may attach an inflation sector
to the model and assume tiny couplings between the inflation sector and
the Higgs sector, which is, at least, technically natural.

For the model to be valid up to the Planck scale, Landau poles should
not appear during the renormalization group (RG) evolution and the
lifetime of the EW vacuum should be long enough. We refer to these two
conditions as the high scale validity. Similar discussions can be found
in the context of THDMs
\cite{Branchina:2018qlf,Krauss:2018thf,Basler:2017nzu,Chakrabarty:2017qkh,Chakrabarty:2016smc,Cacchio:2016qyh,Bagnaschi:2015pwa,Chowdhury:2015yja,Ferreira:2015rha,Das:2015mwa,Chakrabarty:2014aya,Grzadkowski:2013rza,Shu:2013uua}. As
we will see and as found in the previous studies, these conditions are
complementary and become very restrictive if combined. Thus, the model
becomes more predictive and it is important to determine the allowed
parameter space precisely.

The lifetime of the EW vacuum is estimated by the bubble nucleation
rate \cite{Coleman:1977py,Callan:1977pt}, which has a form of
\begin{equation}
 \gamma=\mathcal Ae^{-\mathcal B},
\end{equation}
where $\mathcal B$ is the Euclidean action of the so-called the bounce,
and $\mathcal A$ represents quantum corrections to $\mathcal B$ having
mass dimension four. In many papers, $\mathcal A$ is assumed to lie
around the typical scale of the problem, but it has been pointed out
\cite{Endo:2015ixx} that such an estimation leads to theoretical
uncertainty of $e^{-\mathcal B\times \mathcal O(10\%)}$ in the
nucleation rate. As we will see later, it can become comparable with the
uncertainties coming from those of the top mass and the strong
coupling. Thus, it is important to calculate both of $\mathcal A$ and
$\mathcal B$ to get a precise vacuum decay rate.

The one-loop calculation of $\mathcal A$ for the SM was first calculated
in \cite{Isidori:2001bm}. Since the treatment of the gauge zero mode had
not been known at that time, the calculation was not complete. Recently,
the correct treatment has been found \cite{Endo:2017tsz} and the
one-loop calculation for the SM has been completed
\cite{Andreassen:2017rzq,Chigusa:2017dux,Chigusa:2018uuj}. In addition,
the analytic expression for $\mathcal A$ at the one-loop level has become
available \cite{Andreassen:2017rzq,Chigusa:2018uuj} for an approximately
scale invariant theory.  Since they are applicable to the case where the
bounce is composed of a single field, we extend them to a multi-field
case in this paper. Differently from the single-field case, there can be
more than one unstable directions and there can appear an additional
zero mode due to a global symmetry breaking. In addition, the
electromagnetic $U(1)$ symmetry can also be broken spontaneously.

Before the analysis of the high scale validity, we impose the
constraints from flavor observables, perturbative unitarity, oblique
parameters and collider searches. For the constraints from flavor
observables, we obtain the $95\%$ exclusion limit in Appendix
\ref{apx_flavor} using the recent experimental values.

We determine the allowed parameter space by utilizing the Monte Carlo
method. We show how much the high scale validity narrows down the
parameter space and discuss the implications on the Higgs couplings and
the Higgs mass splittings.

This paper is organized as follows. In Section \ref{sec_model}, we
briefly explain the DFSZ axion model. Section \ref{sec_decay_rate} is
devoted to the details of the analysis on the bubble nucleation rate
for the multi-field case. Then, in Section \ref{sec_low_energy}, we
discuss the low energy constraints. In Section \ref{sec_high_scale}, we
execute numerical analysis and discuss the consequence of the high scale
validity. Finally, we summarize in Section \ref{sec_summary}.

\section{DFSZ Axion Model}\label{sec_model}
In this section, we briefly review the DFSZ axion model
\cite{Zhitnitsky:1980tq,Dine:1981rt}. The scalar sector consists of two
Higgs doublets, $H_1$ and $H_2$, and a SM singlet complex scalar,
$\Phi$. We choose the PQ charges of $H_1$, $H_2$ and $\Phi$ to be $x_1$,
$x_2$ and $(x_2-x_1)/2$, respectively. Here, we assume $x_1\neq x_2$ so
that $\Phi$ has a non-zero PQ charge.

The general scalar potential is given by
\begin{align}
 V(H_1,H_2,\Phi)&=\tilde m_1^2H_1^\dagger H_1+\tilde m_2^2H_2^\dagger H_2+\frac{\lambda_1}{2}(H_1^\dagger H_1)^2+\frac{\lambda_2}{2}(H_2^\dagger H_2)^2\nonumber\\
&\hspace{3ex}+\lambda_3(H_1^\dagger H_1)(H_2^\dagger H_2)+\lambda_4(H_1^\dagger H_2)(H_2^\dagger H_1)\nonumber\\
&\hspace{3ex}+\tilde\lambda_\Phi(|\Phi|^2-v_\Phi^2)^2\nonumber\\
&\hspace{3ex}+ |\Phi|^2(\tilde \kappa_1H_1^\dagger H_1+\tilde \kappa_2H_2^\dagger H_2)-(\tilde \kappa_3\Phi^2H_2^\dagger H_1+h.c.),\label{eq_dfsz_pot}
\end{align}
where $v_\Phi^2$, $\tilde m_i^2$'s, $\lambda_i$'s, $\tilde\lambda_\Phi$
and $\tilde \kappa_i$'s are constants. We assume $\tilde\lambda_\Phi$ is
moderate so that the VEV of $\Phi$ is not affected by those of $H_1$ and
$H_2$.

We assume the classical scale invariance and set $\tilde m_1^2$ and
$\tilde m_2^2$ to zero at the Planck scale. Then, the Higgs mass terms
are assumed to be generated through the PQ symmetry breaking. In order
to obtain the EW scale, $\tilde \kappa_i$'s should be very small since
$\Phi$ has to develop a huge vacuum expectation value (VEV) to avoid the
constraints on the axion decay constant, $10^9~{\rm GeV}\lesssim
f_a\lesssim10^{12}~{\rm GeV}$ \cite{Peccei:2006as,Turner:1989vc}. Due to
the smallness of $\tilde \kappa_i$'s, $\Phi$ decouples from the Higgs
sector and the potential reduces to
\begin{align}
 V_{\rm THDM}&=m_1^2H_1^\dagger H_1+m_2^2H_2^\dagger H_2-(m_3^2H_2^\dagger H_1 +h.c.)+\frac{\lambda_1}{2}(H_1^\dagger H_1)^2\nonumber\\
&\hspace{3ex}+\frac{\lambda_2}{2}(H_2^\dagger H_2)^2+\lambda_3(H_1^\dagger H_1)(H_2^\dagger H_2)+\lambda_4(H_1^\dagger H_2)(H_2^\dagger H_1),
\end{align}
where
\begin{align}
 m_1^2&=\tilde \kappa_1v_\Phi^2,\\
 m_2^2&=\tilde \kappa_2v_\Phi^2,\\
 m_3^2&=|\tilde \kappa_3|v_\Phi^2.
\end{align}
Here, we took $m_3^2$ to be real and positive by the redefinition of the
phase of $H_1$. Notice that PQ violating quartic couplings can be
generated after the PQ symmetry breaking, but they are suppressed by
$\tilde \kappa_{i}$'s and hence are negligible.

With the PQ charge assignment shown in Table \ref{tbl_PQ}, the Higgs
doublets couple to the SM fermions as
\begin{align}
 \mathcal L_{\rm Yukawa}=-y_U\bar Q\tilde H_2U-y_D\bar QH_1D-y_E\bar LH_1E+h.c.,
\end{align}
with
\begin{equation}
 \tilde H_2=i\sigma_2H_2^*.
\end{equation}
Here, $i\sigma_2$ is the completely anti-symmetric matrix and $Q$, $L$,
$U$, $D$ and $E$ represent the left quark doublets, the left lepton
doublets, the up-type quarks, the down-type quarks and the charged
leptons in the SM, respectively.  The model is thus regarded as the
type-II THDM with a limited number of Higgs quartic couplings.

\begin{table}[t]
\begin{center}
  \begin{tabular}{cccccccc}
  \hline\hline
   \multicolumn{8}{c}{PQ Charge Assignment}\\
  \hline\hline
  $H_1$&$H_2$ &$\Phi$ &$Q$ &$L$ &$U$ &$D$ &$E$ \\
  \hline
  $x_1$&$x_2$&$\frac{x_2-x_1}{2}$&$0$&$0$&$x_2$&$-x_1$&$-x_1$\\
  \hline\hline\hline
 \end{tabular}
 \caption{Assignment of the PQ charge in the DFSZ axion model.}
\label{tbl_PQ}
\end{center}
\end{table}

Let us define the mass eigenstates and the mixing angles.  We expand the
Higgs fields as
\begin{equation}
 H_j=
\begin{pmatrix}
 \omega_j^+\\
 (v_j+h_j-i\zeta_j)/\sqrt{2}
\end{pmatrix},
\end{equation}
with
\begin{equation}
 \begin{pmatrix}
  h_1\\
  h_2
 \end{pmatrix}=R(\alpha)
\begin{pmatrix}
 H\\
 h
\end{pmatrix},~
\begin{pmatrix}
  \zeta_1\\
  \zeta_2
 \end{pmatrix}=R(\beta)
\begin{pmatrix}
 G^0\\
 A
\end{pmatrix},~
\begin{pmatrix}
  \omega_1^+\\
  \omega_2^+
 \end{pmatrix}=R(\beta)
\begin{pmatrix}
 G^+\\
 H^+
\end{pmatrix},
\end{equation}
where $v_i$'s are the VEVs of the Higgs fields, $\tan\beta=v_2/v_1$ and
\begin{equation}
 R(\theta)=
\begin{pmatrix}
 \cos\theta&-\sin\theta\\
 \sin\theta&\cos\theta 
\end{pmatrix}.
\end{equation}
Here, $h$ is the $125~{\rm GeV}$ Higgs boson, $H$ is the additional
CP-even Higgs boson, $A$ is the CP-odd Higgs boson, $H^+$ is the charged
Higgs boson, and $G^0$ and $G^+$ are the would-be Nambu-Goldstone
bosons. The SM-like limit for $h$ is given by $\beta-\alpha\to\pi/2$.
\section{Vacuum Decay Rate}\label{sec_decay_rate}
Since the quantum corrections to the effective potential depend on the
VEVs of the Higgs fields, the shape of the effective potential is
non-trivial at large Higgs VEVs. When there is a deeper vacuum or the
effective potential is unbounded from below, the EW vacuum is not
absolutely stable and decays through quantum tunneling. Even in such a
case, we can live in the meta-stable vacuum if it has a much longer
lifetime than the age of the Universe. In this section, we discuss the
precise determination of vacuum decay rates for the DFSZ axion model.

\subsection{Formulation}
Recently, the analytic formulas for the prefactor, $\mathcal A$, at the
one-loop level have been derived
\cite{Andreassen:2017rzq,Chigusa:2018uuj}, which are applicable to the
case where the theory is approximately scale invariant and the bounce
consists of a single field.  In the following, we extend their results
to the case where the bounce consists of more than one fields.

Since the PQ-breaking sector couples to the THDM sector very weakly, the
vacuum decay rate can be calculated independently of the PQ-breaking
sector, {\it i.e.}  the decay path, the RG running or the calculation of
$\mathcal A$ is not affected by the PQ-breaking sector\footnote{If the
potential of the PQ field itself is unstable, we need to calculate the
vacuum decay rate in the PQ sector and add it to that in the THDM
sector. In this paper, we assume the stable potential of the PQ field
given in Eq.~\eqref{eq_dfsz_pot}.}. Notice that even when the field
value of $H_1$ or $H_2$ becomes much larger than the PQ-breaking scale,
$\Phi$ is almost constant during the tunneling. This is because the
typical size of a bounce, {\it i.e.} $\bar
R\simeq1/\sqrt{|H_1(0)|^2+|H_2(0)|^2}$, is too small.  Here, $H_i(0)$'s
are the field values at the center of the bounce. For example, let us
assume that $\Phi$ obtains a negative mass squared, $m_\Phi^2<0$, during the
tunneling. Then, the displacement of $\Phi$ is roughly estimated as
$v_\Phi (e^{\sqrt{|m_\Phi^2|}\bar R}-1)$, which is negligible since
$|m_\Phi^2|\ll 1/\bar R^2$.

Since the field value at the true vacuum is typically much larger than
the EW scale\footnote{The another vacuum may be close to the EW vacuum,
which happens when the low energy potential already has an instability
and the RG running cures it above the EW scale. We will put an IR cut-off
on the size of the bounce to avoid such a situation.}, the Higgs potential is
approximately given by
\begin{align}
 V_{\rm THDM}&\simeq\frac{\lambda_1}{2}(H_1^\dagger H_1)^2+\frac{\lambda_2}{2}(H_2^\dagger H_2)^2+\lambda_3(H_1^\dagger H_1)(H_2^\dagger H_2)+\lambda_4(H_1^\dagger H_2)(H_2^\dagger H_1).\label{eq_pot4}
\end{align}
For the moment, we fix the renormalization scale and will discuss the
running effect later.

The bounce is a solution to the Euclidean equations of motion that are
given by
\begin{align}
 \frac{d^2H_1^i}{dr^2}+\frac{3}{r}\frac{dH_1^i}{dr}&=\frac{\partial V_{\rm THDM}}{\partial H_1^{i*}},\\
 \frac{d^2H_2^i}{dr^2}+\frac{3}{r}\frac{dH_2^i}{dr}&=\frac{\partial V_{\rm THDM}}{\partial H_2^{i*}},
\end{align}
with boundary conditions,
\begin{equation}
 \frac{dH_1^i}{dr}(0)=\frac{dH_2^i}{dr}(0)=0,~H_1^i(\infty)=H_2^i(\infty)=0,
\end{equation}
and their complex conjugates. Here, $r$ is the radius from the center of
the bubble and $i=1,2$ labels the components of the doublet.  Without
loss of generality\footnote{We work in the Fermi gauge as in
\cite{Chigusa:2018uuj} and we pick up one representative element.}, we
parameterize the Higgs fields as
\begin{equation}
 H_1=
\frac{1}{\sqrt{2}}\begin{pmatrix}
 0\\
 \phi\cos\Omega
\end{pmatrix},~H_2=\frac{1}{\sqrt{2}}
e^{i(\sigma_1\theta_1+\sigma_2\theta_2)}e^{i\sigma_3\theta_3}\begin{pmatrix}
 0\\
 \phi\sin\Omega
\end{pmatrix},
\end{equation}
where $\sigma_i$'s are the Pauli matrices. 
Then, the potential is expressed as
\begin{equation}
 V_{\rm THDM}\simeq \frac{\lambda_\phi(\Omega,\Theta)}{4}\phi^4,
\end{equation}
where
\begin{align}
 \lambda_\phi(\Omega,\Theta)&=\frac{1}{2}\left[\lambda_1\cos^4\Omega+\lambda_2\sin^4\Omega+2(\lambda_3+\lambda_4\cos^2\Theta)\sin^2\Omega\cos^2\Omega\right],\label{eq_lambda_phi}\\
 \Theta&=\sqrt{\theta_1^2+\theta_2^2}.
\end{align}

In Appendix \ref{apx_straight}, we show that $\Omega$ and $\Theta$ are
constant\footnote{Quantum corrections to the bounce may depend on
$\Omega$ or $\Theta$. However, they result in two- or higher-loop
corrections to a vacuum decay rate since the bounce is a saddle point
of the action.}. Then, the equations of motion reduce to\footnote{Since
the potential is independent of $\theta_3$, there exist an infinite
number of bounce solutions and a zero mode appears in the calculation of
the functional determinant. We follow \cite{Chigusa:2018uuj} for the
treatment of the zero mode.}
\begin{align}
 \frac{\partial \lambda_\phi}{\partial\Theta}&=0,\label{eq_theta}\\
 \frac{\partial \lambda_\phi}{\partial\Omega}&=0,\label{eq_omega}\\
 \frac{d^2\phi}{dr^2}+\frac{3}{r}\frac{d\phi}{dr}&=\lambda_\phi\phi^3\label{eq_reduced_eom},
\end{align}
with boundary conditions
\begin{equation}
 \frac{d\phi}{dr}(0)=0,~\phi(\infty)=0.\label{eq_bdy_cond}
\end{equation}

From Eq.~\eqref{eq_lambda_phi}, we can see that a minimum of
$\lambda_\phi$ satisfies $\cos^2\Theta=0$ for $\lambda_4\geq0$, and
$\cos^2\Theta=1$ for $\lambda_4<0$. Then, from Eq.~\eqref{eq_omega}, we
get the following solutions;
\begin{align}
 (a)&~\Omega=0,~\lambda_\phi=\frac12\lambda_1,\\
 (b)&~\Omega=\frac{\pi}{2},~\lambda_\phi=\frac12\lambda_2,\\
 (c)&~\tan^2\Omega=\frac{\lambda_1-\bar{\lambda}}{\lambda_2-\bar{\lambda}},~\lambda_\phi=\frac12\frac{\lambda_1\lambda_2-\bar{\lambda}^2}{\lambda_1+\lambda_2-2\bar{\lambda}},
\end{align}
where
\begin{equation}
 \bar{\lambda}=\min(\lambda_3,\lambda_3+\lambda_4).
\end{equation}
Notice that $(c)$ exists only when
$(\lambda_1-\bar{\lambda})/(\lambda_2-\bar{\lambda})>0$.

If $\lambda_\phi<0$, the solution to Eqs.~\eqref{eq_reduced_eom} and
\eqref{eq_bdy_cond} is given by
\begin{equation}
 \phi(r)=\sqrt{\frac{8}{|\lambda_\phi|}}\frac{R}{R^2+r^2},
\end{equation}
which gives
\begin{equation}
 \mathcal B=\frac{8\pi^2}{3|\lambda_\phi|},
\end{equation}
with $R$ being a free parameter that fixes the radius of the bounce.
Notice that $\mathcal B$ is independent of $R$, which is due to the
(approximate) classical scale invariance.

Since all the possible bounces contribute to the vacuum decay rate, the
total vacuum decay rate is expressed as
\begin{equation}
 \gamma=\sum_{\lambda_\phi}\int dR\frac{d\gamma}{dR},
\end{equation}
where $\lambda_\phi$ is summed over its minima with $\lambda_\phi<0$.
Now, the problem is reduced to the single field case for each
$\lambda_\phi$ and we can use the one-loop results of
\cite{Andreassen:2017rzq,Chigusa:2017dux,Chigusa:2018uuj}. The details
are in Appendix \ref{apx_oneloop}.

Let us discuss the convergence of the $R$ integral. From the dimensional
analysis and the renormalization scale independence of the vacuum decay
rate, the $R$-dependence of the integrand can be determined as
\begin{equation}
 \frac{d\gamma}{dR}\propto R^{-5}(\mu R)^{-\frac{8\pi^2\beta_{\lambda_\phi}^{(1)}}{3\lambda_\phi^2}},
\end{equation}
at the one-loop level. Here, $\mu$ is the renormalization scale and
$\beta_{\lambda_\phi}^{(1)}$ is the one-loop beta function for
$\lambda_\phi$. Thus, if we integrate it over $R\in(0,\infty)$, the
integration does not converge.  However, as discussed in
\cite{Andreassen:2017rzq,Chigusa:2017dux,Chigusa:2018uuj}, the result
can be convergent if we include higher-loop corrections. Although it is
very difficult to calculate them, their $R$-dependence is completely
determined by the beta functions and we can sum up the logarithmic
corrections by taking $\mu\sim1/R$ for each bounce with radius $R$ (for
detailed discussion see \cite{Chigusa:2018uuj}). If there exists a
minimum of the effective action, it dominates the $R$ integral and the
result is convergent. 

Independently of the convergence of the $R$ integral, we use cut-offs
for the $R$ integral for the following reasons.  First, we need an IR
cut-off because we have ignored the dimensionful couplings\footnote{The
effect of the mass term of the bounce field at the false vacuum, $m^2$,
is discussed in \cite{Andreassen:2017rzq} and is shown to be suppressed
by $R^2m^2$. We will discuss the cut-off dependence later.}. Second, we
need a UV cut-off because we do not consider gravitational
corrections. Thus, we set the integration region as\footnote{We will
discuss the cut-off dependence later.}
\begin{equation}
 \gamma=\sum_{\lambda_\phi}\int_{1/M_{\rm Pl}}^{1/(10~{\rm TeV})}dR\frac{d\gamma}{dR},\label{eq_cutoff1}
\end{equation}
where $M_{\rm Pl}$ is the reduced Planck scale.  We also impose the same
limits on the field value of the bounce as
\begin{equation}
 10~{\rm TeV}\lesssim\phi(0)=\sqrt{\frac{8}{|\lambda_\phi(\mu)|}}\frac{1}{R}\lesssim M_{\rm Pl},\label{eq_cutoff2}
\end{equation}
when $\lambda_\phi<0$. 

We also exclude the region where the quantum corrections to the action
become larger than $80\%$ of $\mathcal B$ since the perturbative
expansions become unreliable. Such a region appears where $\lambda_\phi$
is very close to zero. 

Since the integrand of the vacuum decay rate is positive definite, these
limits always make the vacuum decay rate small. Thus, what we get with
these limits is a lower bound on the vacuum decay rate and it always
gives a conservative constraint.

The condition for the stability of the EW vacuum is then given by
\begin{equation}
 \gamma\lesssim H_0^4,
\end{equation}
where $H_0\simeq67.66~{\rm (km/s)/Mpc}$ \cite{Aghanim:2018eyx} is the
current Hubble constant.

\subsection{Example}
\begin{figure}[t]
 \begin{center}
  \begin{minipage}{0.37\linewidth}
   \includegraphics[width=\linewidth]{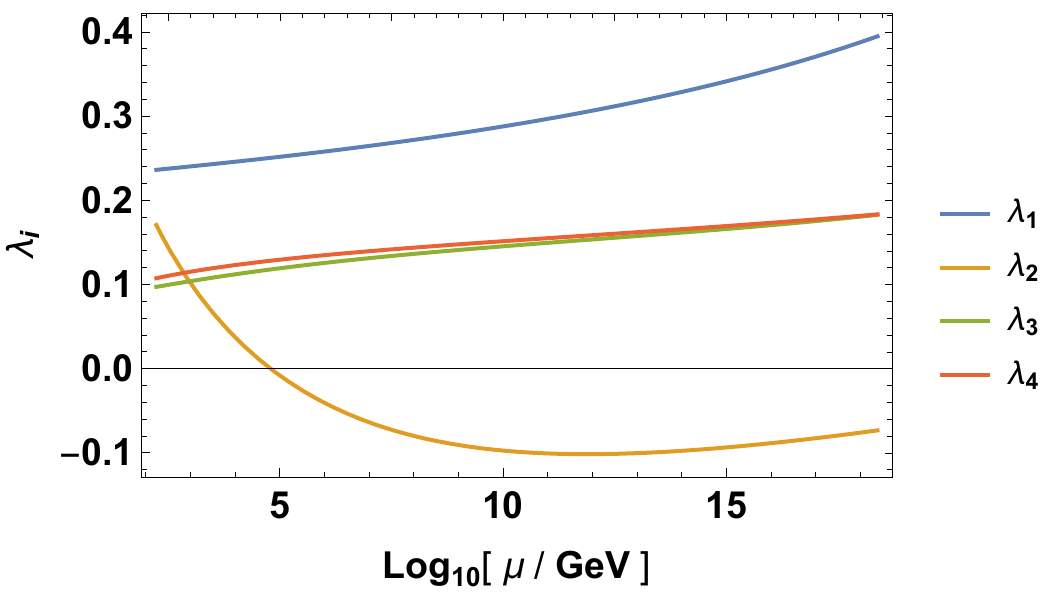}
  \end{minipage}\hspace{2ex}
  \begin{minipage}{0.43\linewidth}
   \includegraphics[width=\linewidth]{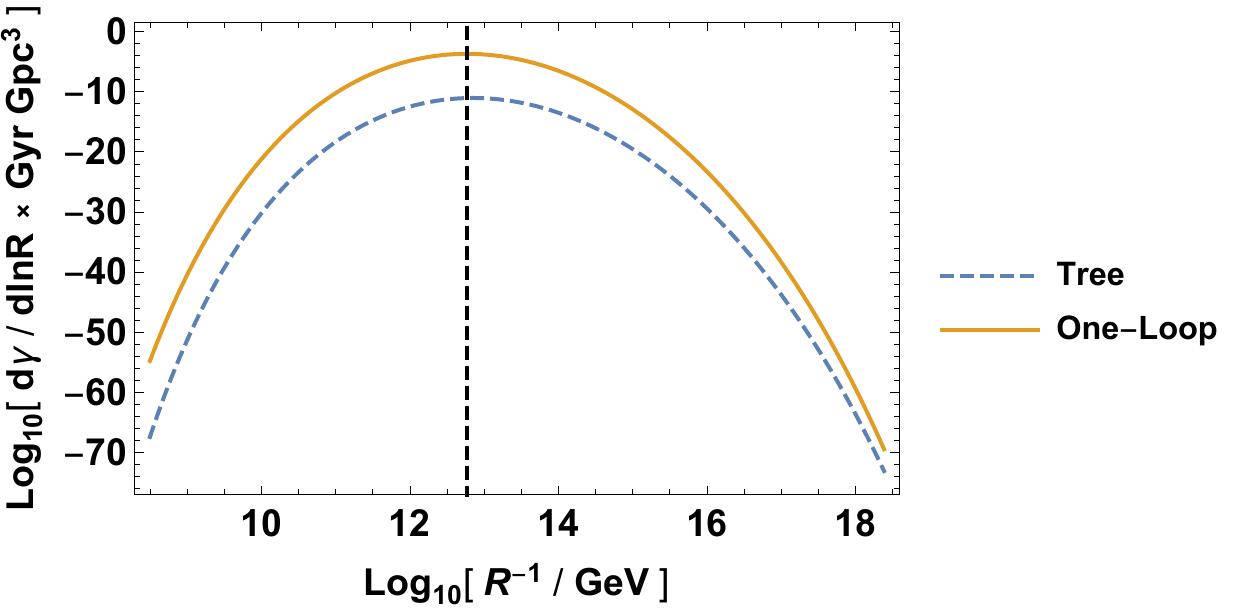}
  \end{minipage}\vspace{3ex}
  \begin{minipage}{0.4\linewidth}
   \includegraphics[width=\linewidth]{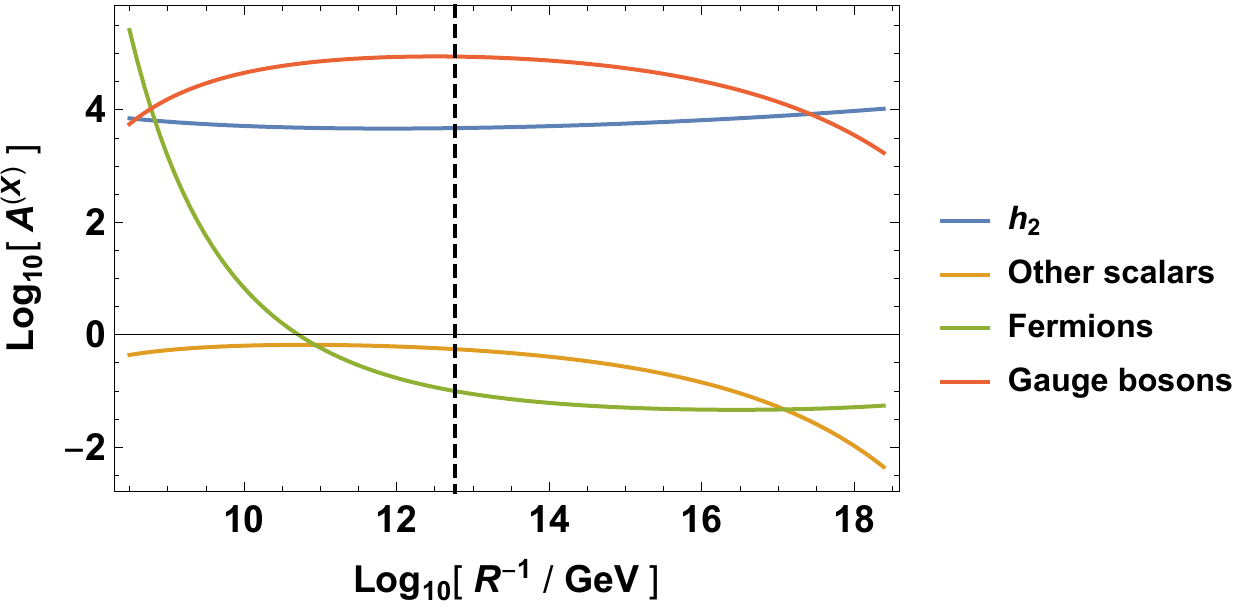}
  \end{minipage}
 \end{center}
\caption{An example of the calculation of a vacuum decay rate. The top
left panel shows the RG evolution of the Higgs quartic couplings. The
top right panel shows the differential vacuum decay rates with and
without the calculation of $\mathcal A$. The bottom panel shows each
quantum correction to the differential vacuum decay rate. The vertical
black dashed line indicates the maximum of the differential vacuum decay
rate.}  \label{fig_example}
\end{figure}

Let us show an example of the calculation. We take
\begin{align}
 &\tan\beta=9,~\cos(\beta-\alpha)=0.0004,\nonumber\\
 &m_H=602.5~{\rm GeV},~m_A=602.5~{\rm GeV},~m_{H^+}=600~{\rm GeV}.
\end{align}
We first calculate the $\msbar$ dimensionless couplings at
renormalization scale $\mu_t=m_t$, where we include the one-loop
corrections and the four-loop QCD corrections. The details are in Appendix
\ref{apx_matching}. Then, we evolve them with the two-loop RG
equations. The result is shown in the top left panel of
Fig.~\ref{fig_example}. In this example, only $\lambda_2$ becomes
negative and contributes to the vacuum decay rate.

Next, we calculate the differential vacuum decay rate, $d\gamma/d(\ln
R)$, for case (b). We take $\mu=1/R$. The result is shown with the solid
line in the top right panel of Fig.~\ref{fig_example}. Integrating it
over $\ln R$, we get
\begin{equation}
 \log_{10}[\gamma\times{\rm Gyr}~{\rm Gpc}^3]=-3.5^{~+21.7~+11.0~+1.4~+0.1}_{~-26.1~-11.8~-1.5~-0.3},
\end{equation}
where the 1st, 2nd, 3rd and 4th errors are those from $m_t$, $\alpha_s$,
$m_h$ and $\mu$, respectively. We use the SM values and uncertainties
given in Table \ref{tbl_param_sm} and $\alpha_s=0.1181(11)$. We estimate
the renormalization scale uncertainty by taking $\mu=2/R$ and
$\mu=1/(2R)$.  With this parameter set, the vacuum decay rate is close
to the upper bound, $\log_{10}[H_0^4\times{\rm Gyr}~{\rm
Gpc}^3]\simeq-3$.

Let us see the difference between the ``tree level'' vacuum decay rate
and our result. For the tree level vacuum decay rate, we adopt
\begin{equation}
 \gamma_{\rm tree}=\max_{R,\lambda_\phi}\left[\frac{1}{R^4}e^{-\frac{8\pi^2}{3|\lambda_\phi|}}\right]_{\mu=R^{-1}},
\end{equation}
where the maximum value is searched in the same region as the
integration region of the one-loop vacuum decay rate. In the top right panel of
Fig.~\ref{fig_example}, we show $R^{-4}\exp[-8\pi^2/(3|\lambda_\phi|)]$
with the dashed line.  We get
\begin{equation}
 \log_{10}[\gamma_{\rm tree}\times{\rm Gyr}~{\rm Gpc}^3]=-11.2.
\end{equation}
Thus, the one-loop calculation enhances the vacuum decay rate by about
$10^{7.7}$.  We show each quantum contribution in the bottom panel of
Fig.~\ref{fig_example}. The vertical black dashed line corresponds to
the maximum of the differential vacuum decay rate. Around the maximum,
the gauge bosons and $h_2$ have positive contributions and the fermions
and the other scalars have negative contributions. The former
contributions are larger than the latter and the positive contribution
remains.

\begin{figure}[t]
 \begin{center}
  \includegraphics[width=0.4\linewidth]{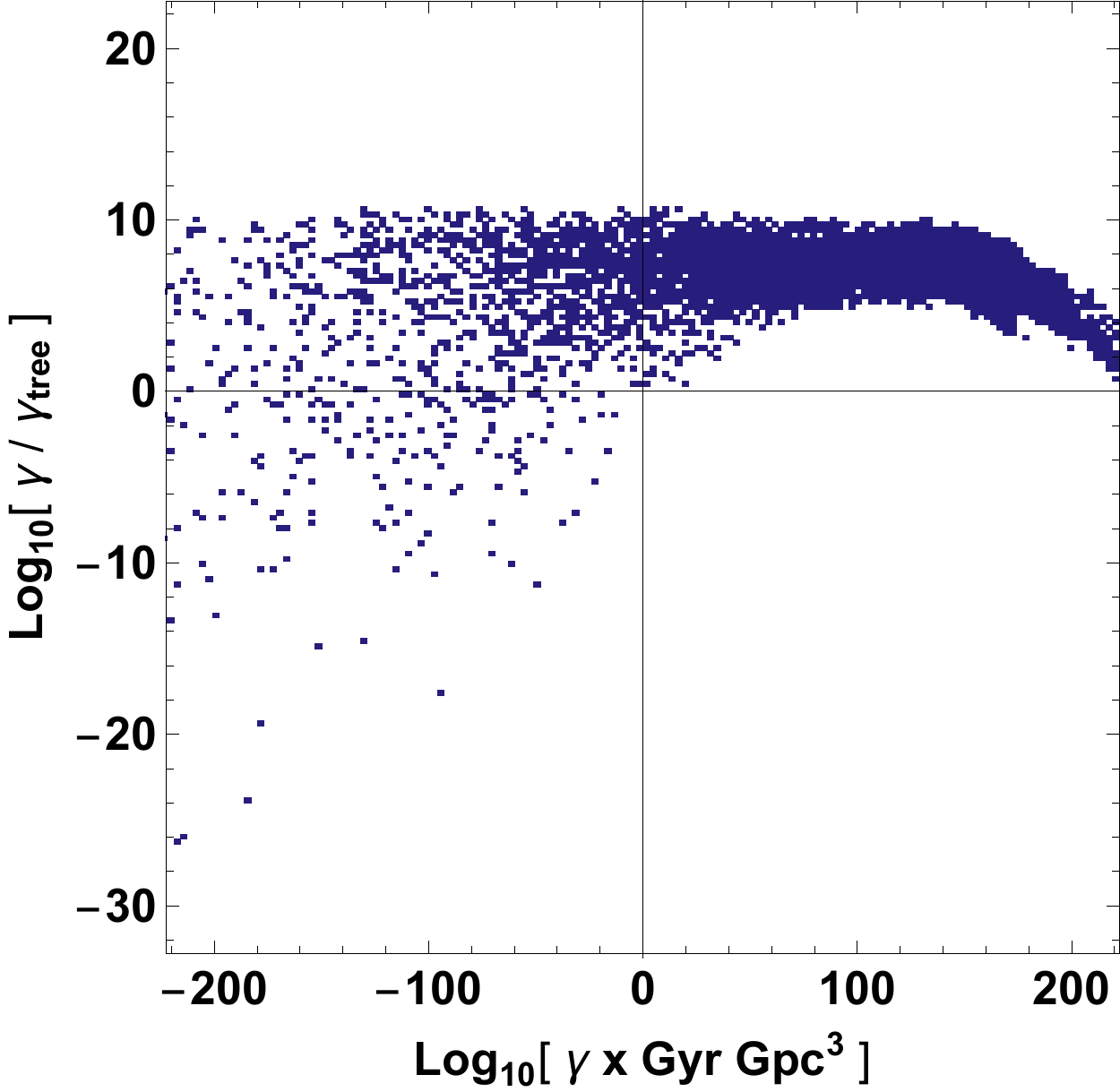}
 \end{center}
\caption{The difference between the tree level vacuum decay rates and
the one-loop level vacuum decay rates. We use the data accumulated for
Fig.~\ref{fig_scat1}. We show only $-200\lesssim\log_{10}[\gamma\times
{\rm Gyr}{\rm Gpc}^3]\lesssim200$.}  \label{fig_tree_1l}
\end{figure}

In Fig.~\ref{fig_tree_1l}, we show the binned plot of the vacuum decay
rates at the tree level and at the one-loop level by using the data
accumulated for Fig.~\ref{fig_scat1}.  We observe that the enhancement
of the vacuum decay rate is generic for $\gamma\gtrsim H_0^4$ and that
it is enhanced at most by $10^{10}$, which is comparable with the
uncertainties from those of the top mass and the strong coupling
constant. For $\gamma\ll H_0^4$, the vacuum decay rate can be either
suppressed or enhanced.

\section{Low Energy Constraints}\label{sec_low_energy}
Before the discussion of the high scale validity, let us discuss the low
energy constraints; flavor observables, perturbative unitarity, oblique
parameters, and collider searches. In this section, we do not consider
the constraints from the signal strengths of the $125~{\rm GeV}$ Higgs
boson since they are the outputs of our analysis.

\subsection{Flavor Observables}
The additional Higgs bosons contribute to flavor observables and the
current strongest constraints for the type-II THDM come from the
branching ratios of $B\to\tau\nu$, $B_s\to\mu\mu$ and $b\to s\gamma$,
and the $B_s-\bar B_s$ mixing as discussed, for example, in
\cite{Arbey:2017gmh,Misiak:2017bgg,Haller:2018nnx}. We obtain the
constraints following the analysis of \cite{Enomoto:2015wbn} with the
current experimental values. The details are given in Appendix
\ref{apx_flavor}.

In Fig.~\ref{fig_flavor}, we plot the $95\%$ exclusion limits on the
$(m_{H^+},\tan\beta)$-plane with assuming $m_H=m_A=m_{H^+}$ and
$\cos(\beta-\alpha)=0$\footnote{These parameters affect only ${\rm
BR}(B_s^0\to\mu^+\mu^-)$. As we will see later, the high scale validity
requires a small $\cos(\beta-\alpha)$ and mass differences. Then, the
result is not so much affected as discussed in
\cite{Enomoto:2015wbn}.}. The white region is allowed and the shaded
regions are excluded by the observables shown on the regions. As we can
see, ${\rm BR}(b\to s\gamma)$ gives the lower bound of
$m_{H^+}\gtrsim580~{\rm GeV}$ almost independently of $\tan\beta$. The
upper bound and the lower bound on $\tan\beta$ are set by ${\rm
BR}(B_s\to\mu\mu)$ and $\Delta M_{B_s}$, respectively. Notice that these
constraints are stronger than the perturbativity limits of
$y_t,y_b\lesssim\sqrt{4\pi}$. The results are consistent with the recent
works\footnote{Since there are choices of input parameters and of the
treatment of theoretical uncertainty, $\mathcal O(10\%)$ difference of
the constraints is acceptable.}
\cite{Arbey:2017gmh,Misiak:2017bgg,Haller:2018nnx}.

\begin{figure}[t]
\begin{center}
 \begin{center}
  \includegraphics[width=0.4\linewidth]{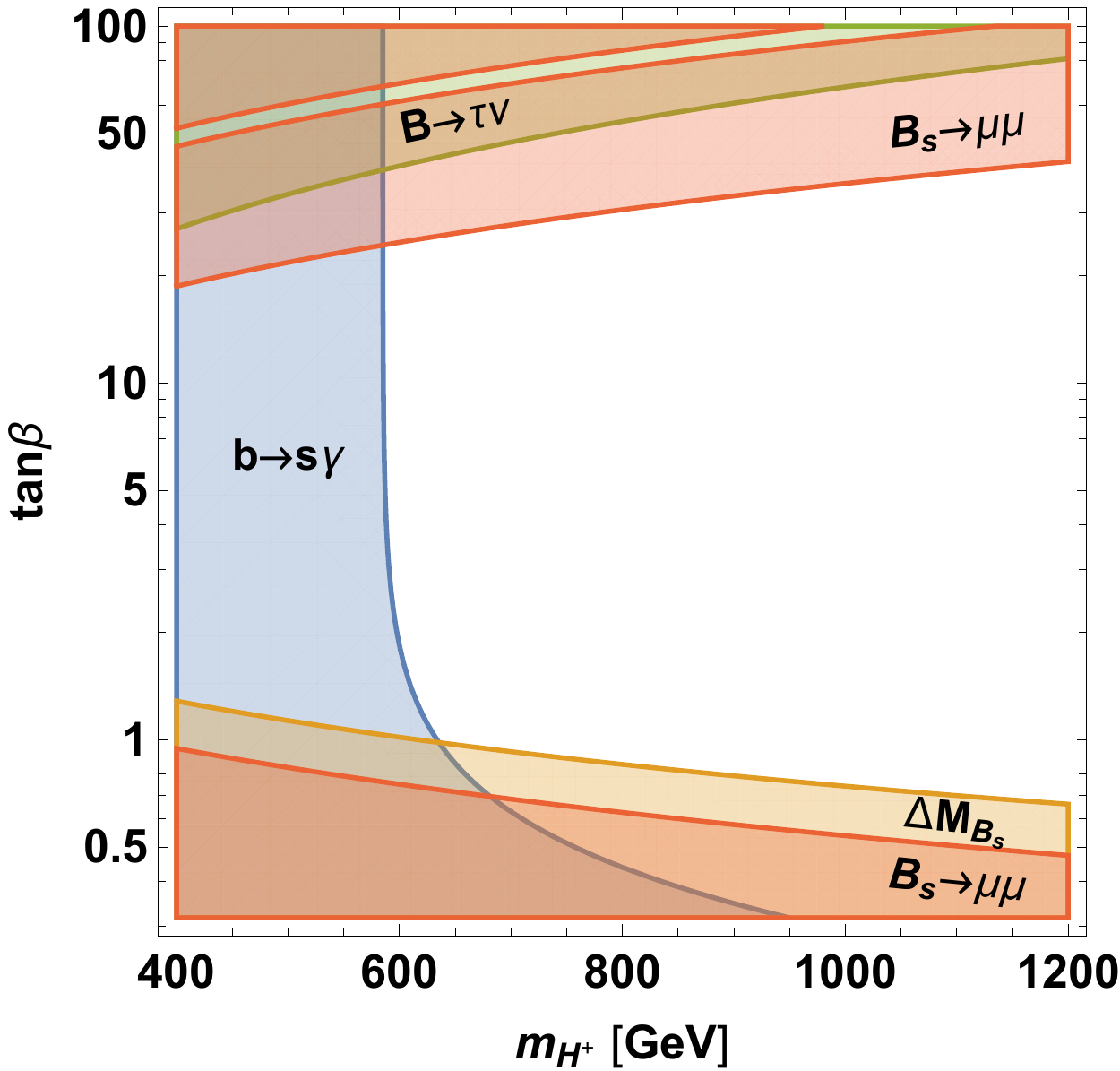}
 \end{center}
\end{center}
\caption{The $95\%$ CL constraints from flavor observables. The blue,
the orange, the red, and the green shaded regions are excluded by ${\rm
BR(b\to s\gamma)}$, $\Delta M_{B_s}$, ${\rm BR}(B_s\to\mu\mu)$, and
${\rm BR}(B\to\tau\nu)$, respectively.}  \label{fig_flavor}
\end{figure}
\subsection{Low Energy Perturbative Unitarity}
For the study of the high scale validity, the perturbative unitarity is
necessary because otherwise all the calculations, including the matching
conditions to the $\msbar$ couplings, become unreliable.

At the tree level, the Higgs quartic couplings are related to the Higgs
masses and mixing as
\begin{align}
 \lambda_1&=\frac{1}{2v^2\cos^2\beta}\left[m_h^2+m_H^2-(1-\cos2\beta)m_A^2+(m_H^2-m_h^2)\cos2\alpha\right],\\
\lambda_2&=\frac{1}{2v^2\sin^2\beta}\left[m_h^2+m_H^2-(1+\cos2\beta)m_A^2-(m_H^2-m_h^2)\cos2\alpha\right],\\
\lambda_3&=\frac{1}{v^2}\left[2m_{H^+}^2-m_A^2+(m_H^2-m_h^2)\frac{\sin2\alpha}{\sin2\beta}\right],\\
 \lambda_4&=\frac{2}{v^2}(m_A^2-m_{H^+}^2).
\end{align}
Then, we impose the condition of the $s$-wave unitarity, which is given
by \cite{Kanemura:1993hm,Akeroyd:2000wc}
\begin{align}
 |\lambda_1|&<8\pi,\label{eq_unitarity_begin}\\
 |\lambda_2|&<8\pi,\\
 |\lambda_3|&<8\pi,\\
 |\lambda_3\pm\lambda_4|&<8\pi,\\
 |\lambda_3+2\lambda_4|&<8\pi,\\
 \left|\frac{1}{2}\left(\lambda_1+\lambda_2\pm\sqrt{(\lambda_1-\lambda_2)^2+4\lambda_4^2}\right)\right|&<8\pi,\\
 \left|\frac{3}{2}\left(\lambda_1+\lambda_2\pm\sqrt{(\lambda_1-\lambda_2)^2+\frac{4}{9}(2\lambda_3+\lambda_4)^2}\right)\right|&<8\pi\label{eq_unitarity_end},
\end{align}
at the tree level\footnote{Precisely speaking, this condition of
perturbative unitarity is valid up to $\mathcal O(1)$
uncertainty. However, it is enough for our purpose of avoiding too large
quartic couplings.}.  Since we will use the same condition to detect
Landau poles later, we refer to the perturbative unitarity with the tree
level matching conditions as ``the low energy (LE) perturbative
unitarity''.

\subsection{Oblique Parameters}
The oblique parameters, especially the $S$-parameter and the
$T$-parameter, are affected by the additional Higgs doublet.  We use the
general formulas for multi-Higgs-doublet models
\cite{Grimus:2007if,Grimus:2008nb} (for the THDM, see
\cite{Eriksson:2009ws}\footnote{There is a typo in the $G(x,y,Q)$
function in \cite{Eriksson:2009ws}. The correct definition is in
\cite{Grimus:2008nb}.}) to calculate these parameters.

The current constraints are given by \cite{Tanabashi:2018oca}
\begin{align}
 S&=0.02\pm0.07,\label{eq_s_par}\\
 T&=0.06\pm0.06,\label{eq_t_par}
\end{align}
with the assumption of $U=0$. The correlation coefficient is
$\rho=0.92$. We adopt $95\%$ exclusion limit on $S$ and $T$, which is
given by
\begin{equation}
 \chi^2_{2\,{\rm dof}}\equiv\frac{1}{1-\rho^2}\left[\frac{(S-S_{\rm cent})^2}{\sigma_S^2}+\frac{(T-T_{\rm cent})^2}{\sigma_T^2}-2\rho\frac{(S-S_{\rm cent})(T-T_{\rm cent})}{\sigma_S\sigma_T}\right]<5.99,
\end{equation}
where $S_{\rm cent}$ and $T_{\rm cent}$ are the central values of $S$
and $T$, respectively.
\subsection{Collider Searches}
New scalar particles have been searched extensively at Tevatron, LEP and
LHC.  We utilize {\tt HiggsBounds}
\cite{Bechtle:2013wla,Bechtle:2013gu,Bechtle:2011sb,Bechtle:2008jh,Bechtle:2015pma}
to check the constraints from the collider searches. For simplicity, we
consider only on-shell decays for non-SM channels. The couplings and the
partial decay widths used in the analysis are summarized in Appendix
\ref{apx_hb}.

\section{High Scale Validity}\label{sec_high_scale}
At an energy scale much higher than the EW scale, the model becomes
classically scale invariant and only the dimensionless couplings become
relevant.  We first match the $\overline{\rm MS}$ couplings at the
one-loop level, where the matching scale is taken to the top mass
scale. For the top and the bottom Yukawa couplings, we also include the
four-loop QCD corrections. Then, we evolve the dimensionless couplings
up to the Plank scale using the two-loop beta functions. In these
calculations, we utilize the public codes of {\tt SARAH}
\cite{Staub:2011dp,Staub:2015kfa}, {\tt FeynArts} \cite{Hahn:2000kx},
{\tt FeynCalc} \cite{Mertig:1990an,Shtabovenko:2016sxi}, and {\tt
RunDec} \cite{Chetyrkin:2000yt,Herren:2017osy}. The details of the
matching conditions are given in Appendix \ref{apx_matching}. Throughout
this analysis, we adopt the central values for the SM inputs, which are
summarized in Table \ref{tbl_param_sm}.

For the model to be valid up to the Planck scale, Landau poles should
not appear during the RG evolution. We adopt the condition of the tree
level perturbative unitarity given in
Eqs.~\eqref{eq_unitarity_begin}-\eqref{eq_unitarity_end} to detect
Landau poles and require that they should be satisfied until the Planck
scale. We refer to this condition as ``the high energy (HE) perturbative
unitarity''. If it is satisfied, we then check the vacuum stability,
where we take $\mu=1/R$.

We reduce the number of free parameters by choosing three slices of
parameter space;
\begin{align}
 (i)~&m_{H^+}=600~{\rm GeV},~1.8<\tan\beta<25,\label{eq_fl_begin}\\
 (ii)~&m_{H^+}=900~{\rm GeV},~0.8<\tan\beta<33,\\
 (iii)~&m_{H^+}=1200~{\rm GeV},~0.65<\tan\beta<40,\label{eq_fl_end}
\end{align}
which satisfy the flavor constraints of Fig.~\ref{fig_flavor}.  Since
the flavor constraints do not depend so much on the other Higgs masses
or $\cos(\beta-\alpha)$ in the region of interest, we do not further
check the flavor constraints to reduce computational complexity. In
addition, we assume $\sin(\beta-\alpha)>0$ in this analysis.

For each slice, we generate random two million data points that satisfy
all of the other low energy constraints, namely, LE perturbative
unitarity, oblique parameters and collider searches. The scattering
range covers all of the parameter space where the LE perturbative
unitarity is satisfied. The details of data generation are in Appendix
\ref{apx_mc}.

In Fig.~\ref{fig_scat1}, we show the binned plots of the allowed data
points. All the colored points satisfy the low energy constraints. In
the upper panels, the large $\tan\beta$ region is excluded by the
$H\to\tau\tau$ channel. For slice (i), the upper and the lower bounds on
$\cos(\beta-\alpha)$ are determined by the constraints on the $H\to VV$
and the $H\to 2h\to 4b$ channels, respectively.  For slices (ii) and
(iii), the upper and the lower bounds on $\cos(\beta-\alpha)$ are mostly
determined by the constraints on the LE perturbative unitarity and the
oblique parameters, respectively. As for the lower panels, the concave
shape is due to the constraint on the oblique parameters and the horns
have the ends due to the other constraints.

Next, the orange and the green points satisfy the HE perturbative
unitarity. The allowed parameter space is reduced especially for slice
(i), but the reduction is not so drastic.

Finally, the green points satisfy the vacuum stability condition. As we
can see, the parameter space is reduced drastically. It is because of
the complementarity of the HE perturbative unitarity and the vacuum
stability. It can be understood from the one-loop beta functions of
$\lambda_1$ and $\lambda_2$, which are given by
\begin{align}
 \beta_{\lambda_1}&=2[6\lambda_1^2+\lambda_3^2+(\lambda_3+\lambda_4)^2]+\frac{3}{4}(g_Y^4+3g_2^4+2g_Y^2g_2^2)\nonumber\\
 &\hspace{3ex}+\lambda_1(12y_b^2+4y_\tau^2-3g_Y^2-9g_2^2)-12y_b^4-4y_\tau^4,\\
 \beta_{\lambda_2}&=2[6\lambda_2^2+\lambda_3^2+(\lambda_3+\lambda_4)^2]+\frac{3}{4}(g_Y^4+3g_2^4+2g_Y^2g_2^2)\nonumber\\
 &\hspace{3ex}+\lambda_2(12y_t^2-3g_Y^2-9g_2^2)-12y_t^4.\label{eq_beta_lambda_2}
\end{align}
Since $y_t,y_b,y_\tau$ and $g_2$ are UV free, $\beta_{\lambda_1}$ and
$\beta_{\lambda_2}$ generically become positive at a high energy
scale. To avoid Landau poles, the quartic couplings should be small
enough. In addition, negative $\lambda_1$ or $\lambda_2$ are preferable
since they delay the appearance of Landau poles. Thus, the potential
easily becomes unstable and a large part of the parameter space is
constrained by the vacuum stability.

\begin{figure}[t]
\begin{center}
\begin{minipage}{0.3\linewidth}
 \includegraphics[width=\linewidth]{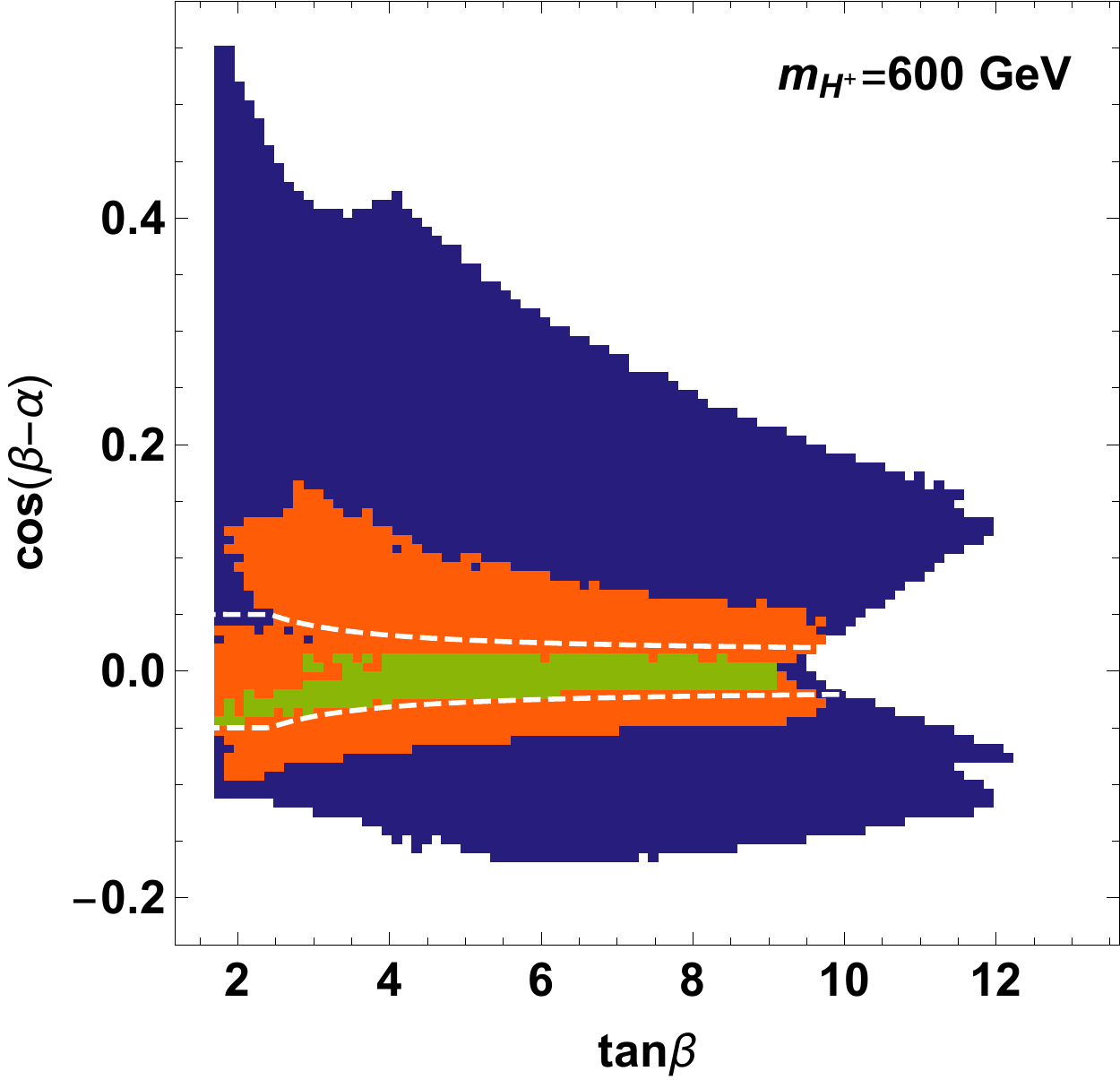}
\end{minipage}\hspace{1.5ex}
\begin{minipage}{0.3\linewidth}
 \includegraphics[width=\linewidth]{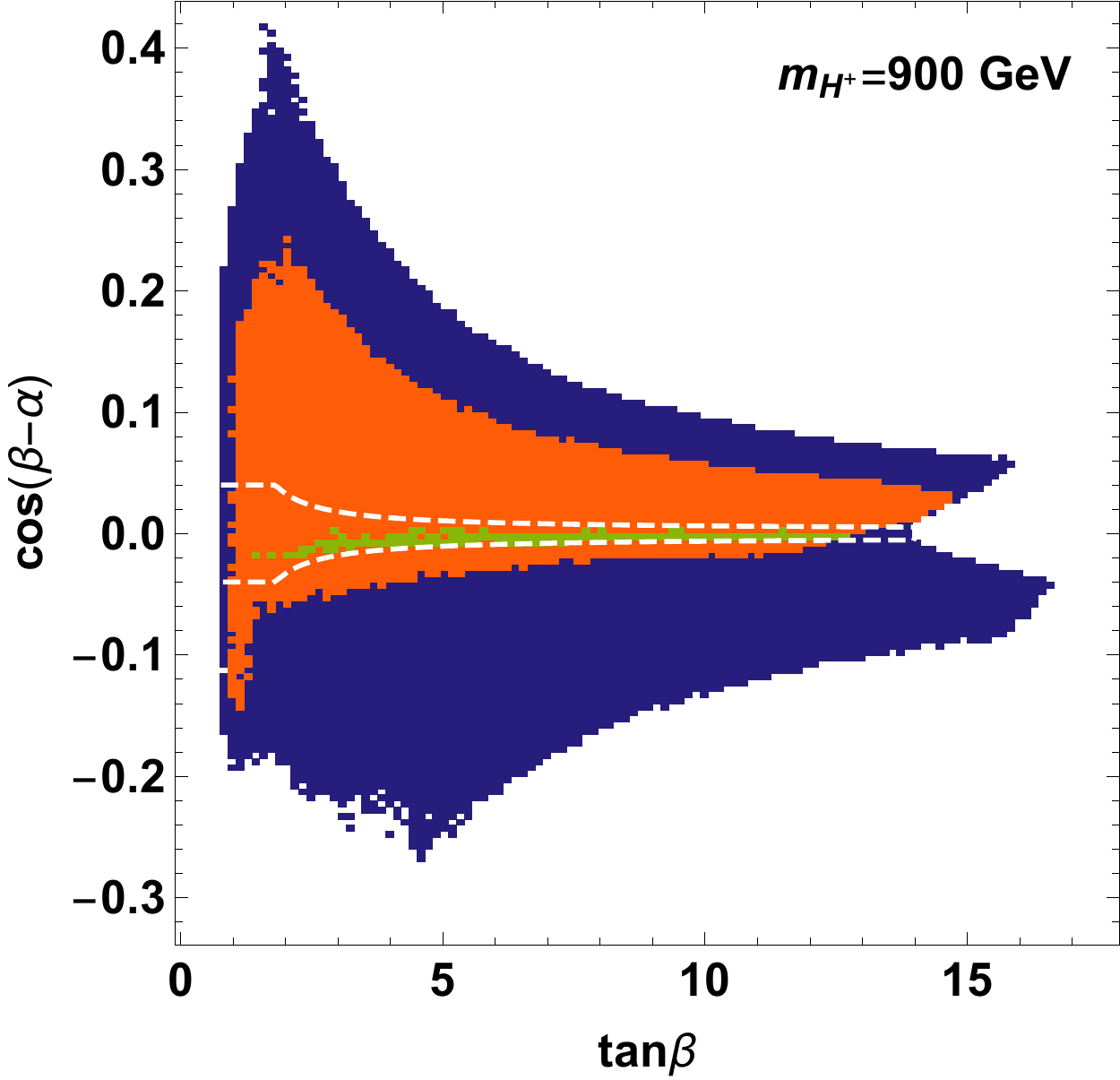}
\end{minipage}\hspace{1.5ex}
\begin{minipage}{0.3\linewidth}
 \includegraphics[width=\linewidth]{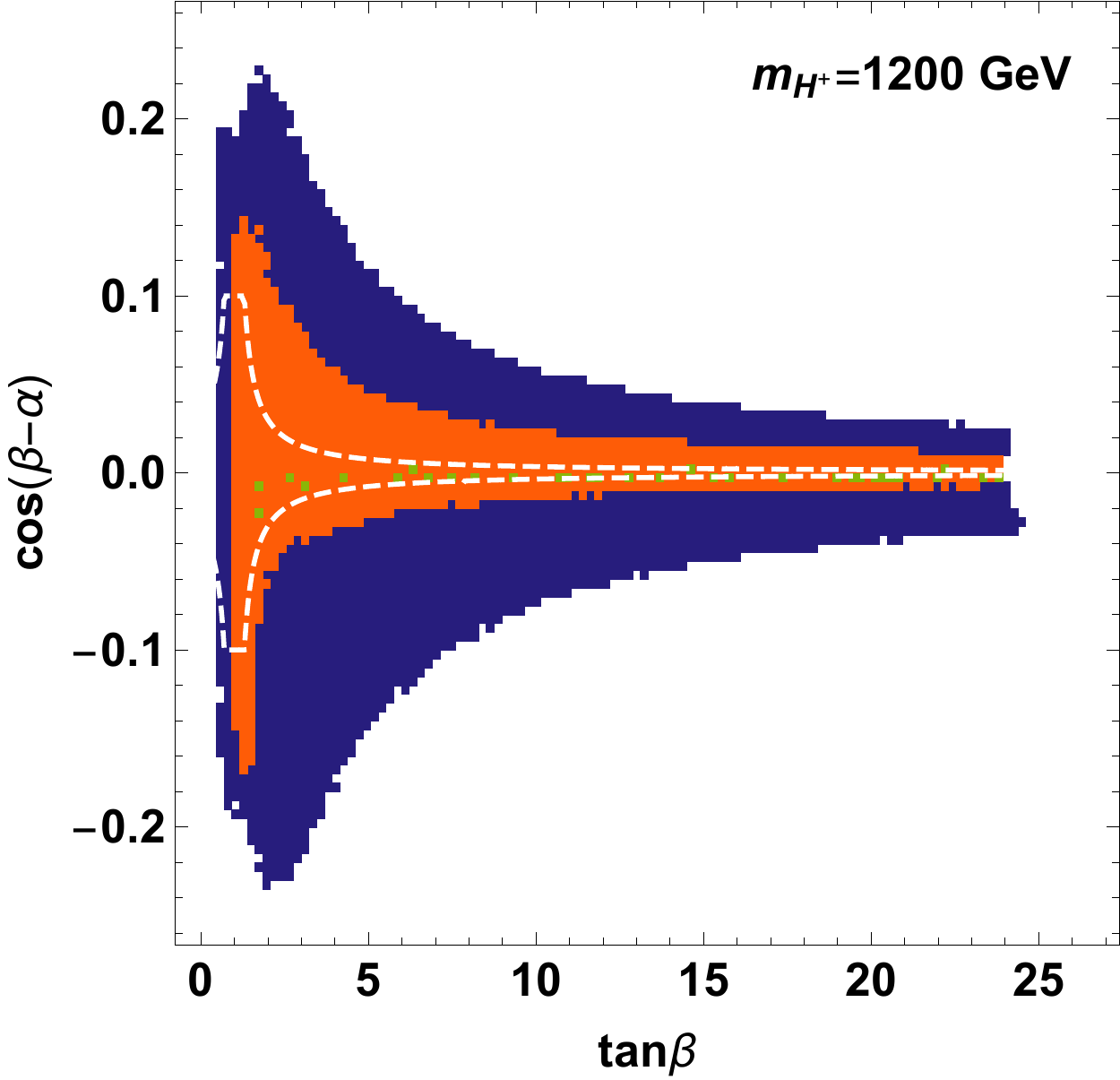}
\end{minipage}
\end{center}
\begin{center}
 \begin{minipage}{0.3\linewidth}
 \includegraphics[width=\linewidth]{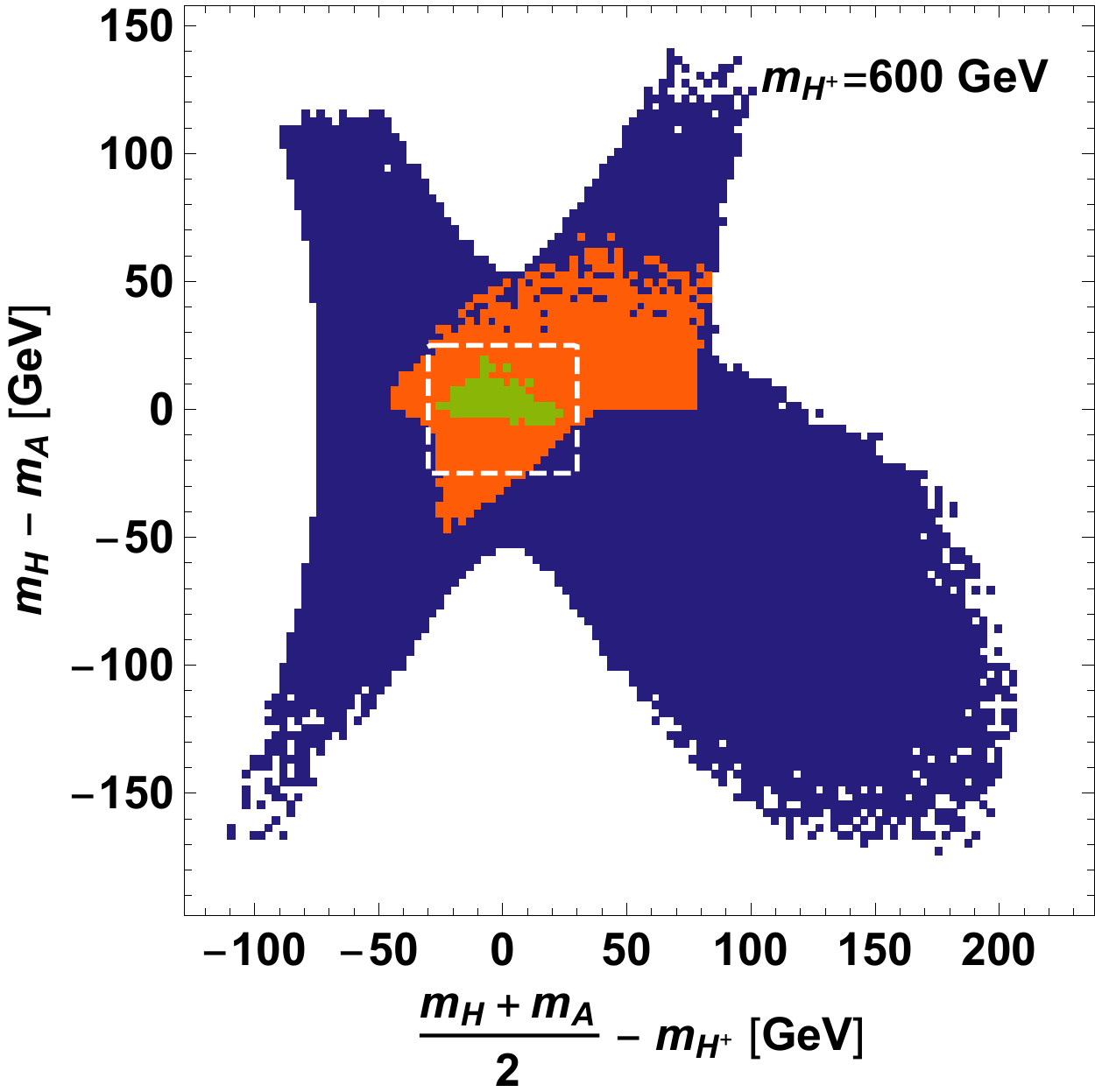}
\end{minipage}\hspace{1.5ex}
\begin{minipage}{0.3\linewidth}
 \includegraphics[width=\linewidth]{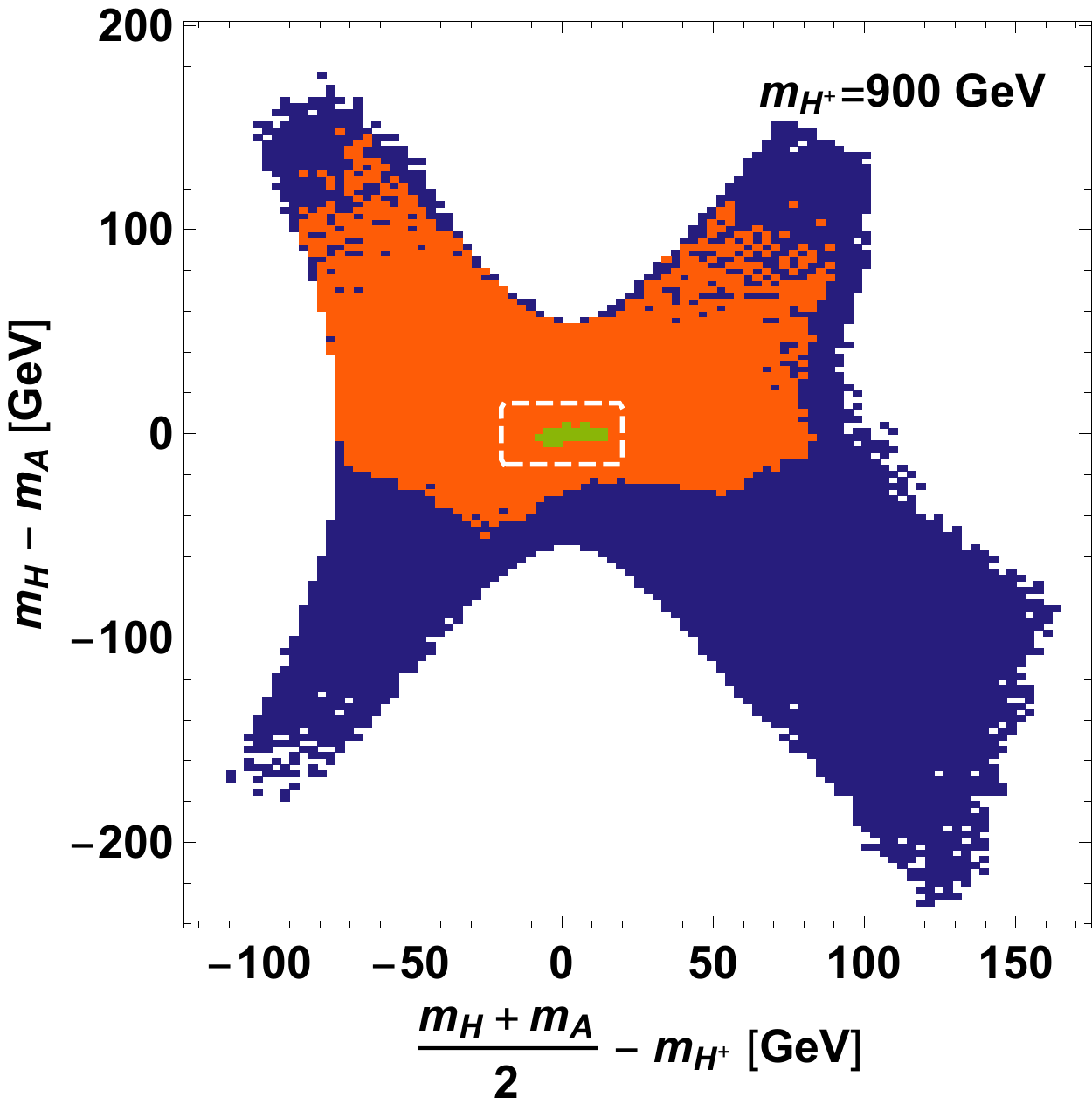}
\end{minipage}\hspace{1.5ex}
\begin{minipage}{0.3\linewidth}
 \includegraphics[width=\linewidth]{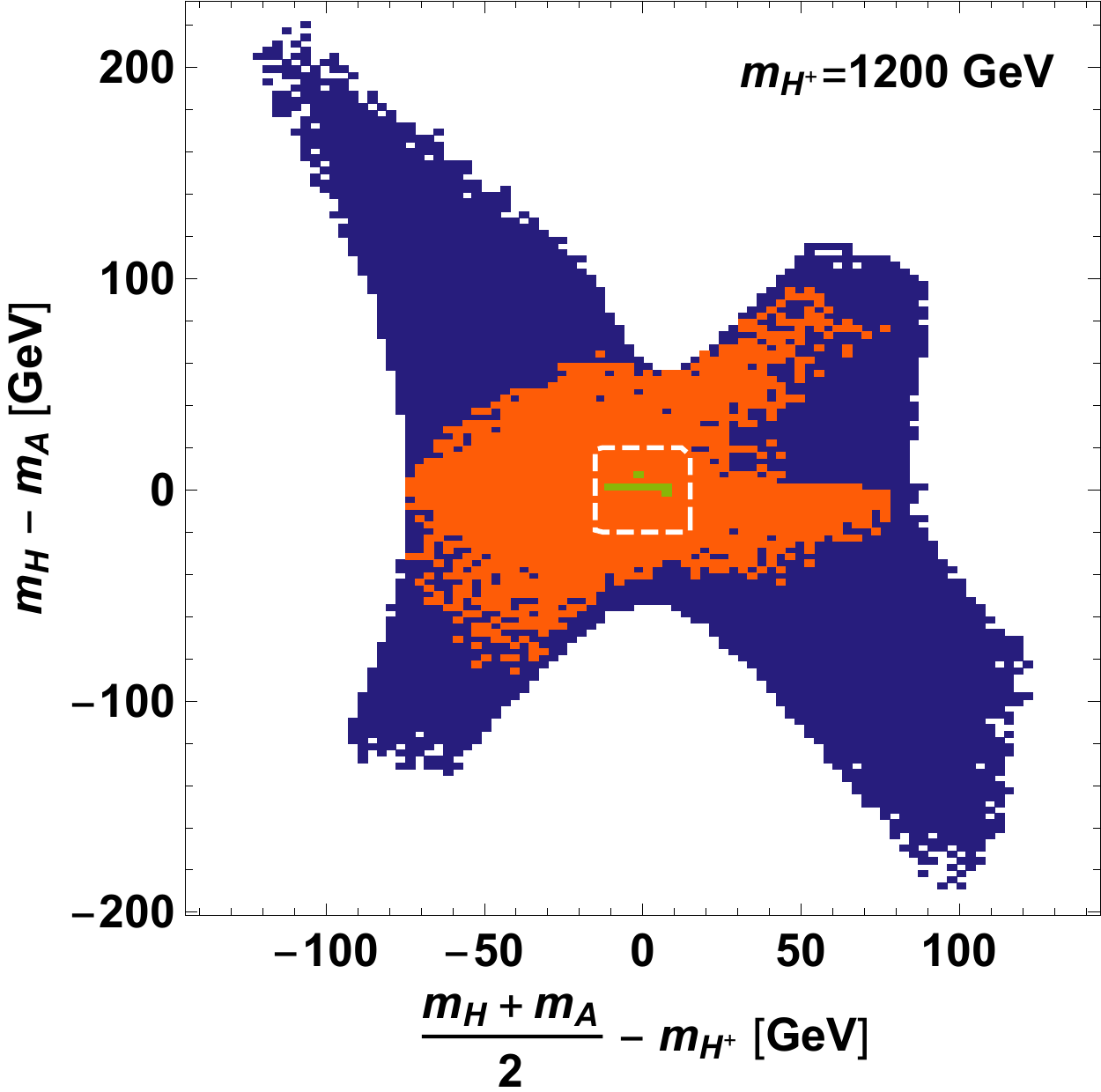}
\end{minipage}
\end{center}
\caption{Binned plots of allowed data points. The left, the middle and
the right panels correspond to the parameter slices of (i), (ii) and
(iii), respectively. All the colored points satisfy the low energy
constraints. The orange and the green points satisfy the perturbative
unitarity conditions until the Planck scale. The green points satisfy
the vacuum stability condition. The regions surrounded by the white
dashed lines are used in Fig~\ref{fig_scat2}. } \label{fig_scat1}
\end{figure}

A similar condition as the vacuum stability is the bounded-from-below
condition, which is given by
\begin{align}
 \lambda_1>0~\&~\lambda_2>0~\&~\left(\frac{\lambda_1\lambda_2-\bar\lambda^2}{\lambda_1+\lambda_2-2\bar\lambda}>0~{\rm or}~\frac{\lambda_1-\bar\lambda}{\lambda_2-\bar\lambda}<0\right).
\end{align}
Here, we regard those couplings as the $\overline{\rm MS}$ couplings at
$\mu=m_t$ and impose it only at low energy. Notice that the condition is
obtained from the discussion of Section \ref{sec_decay_rate} and is
equivalent to that in \cite{Deshpande:1977rw}. We expect that the
combination of the HE perturbative unitarity and the bounded-from-below
condition should give a similar result\footnote{If we impose only the
bounded-from-below condition and the low energy constraints, the allowed
region is as large as the orange region of Fig.~\ref{fig_scat1}.}, which
we will see below.

In Fig.~\ref{fig_scat2}, we pick up the parameter space defined by the
region surrounded by the white dashed lines in Fig.~\ref{fig_scat1} and
prepare additional five million points satisfying all the low energy
constraints for each region. The scattering region is taken so that it
can cover all the green points. The distribution of the new data points
is uniform in the space of $\tan\beta$,
$\cos(\beta-\alpha)/|\cos(\beta-\alpha)|_{\max}$, $m_H-m_A$ and
$(m_H+m_A)/2-m_{H^+}$. Here, $|\cos(\beta-\alpha)|_{\max}$ is the
maximum value of $\cos(\beta-\alpha)$ depending on $\tan\beta$, which is
shown in Fig.~\ref{fig_scat1}. The red points satisfy the
bounded-from-below condition and the HE perturbative unitarity. The
lighter and the darker green points correspond to the green points in
Fig.~\ref{fig_scat1} and are plotted over the red points. Thus, in the
red region appearing in the figure, the potential is stable at low
energy, but always becomes unstable at high energy. The darker green
points satisfy both the vacuum stability and the bounded-from-below
conditions. Thus, in the lighter green region, the potential always
becomes unstable at low energy, but the instability is cured at high
energy. Notice that the vacuum decay rates can be affected by the IR
cut-off for the $R$ integral in the lighter green region.

As we can see from the figure, the bounded-from-below condition has a
similar effect as the vacuum stability condition, but the allowed
regions do not overlap completely. In particular, a large part of the
region with $m_H<m_A$ is excluded by the vacuum stability, where
$\lambda_2$ tends to become negative during the RG evolution. In
addition, a negative $\cos(\beta-\alpha)$ is more favored by the vacuum
stability.

\begin{figure}[t]
\begin{center}
\begin{minipage}{0.3\linewidth}
 \includegraphics[width=\linewidth]{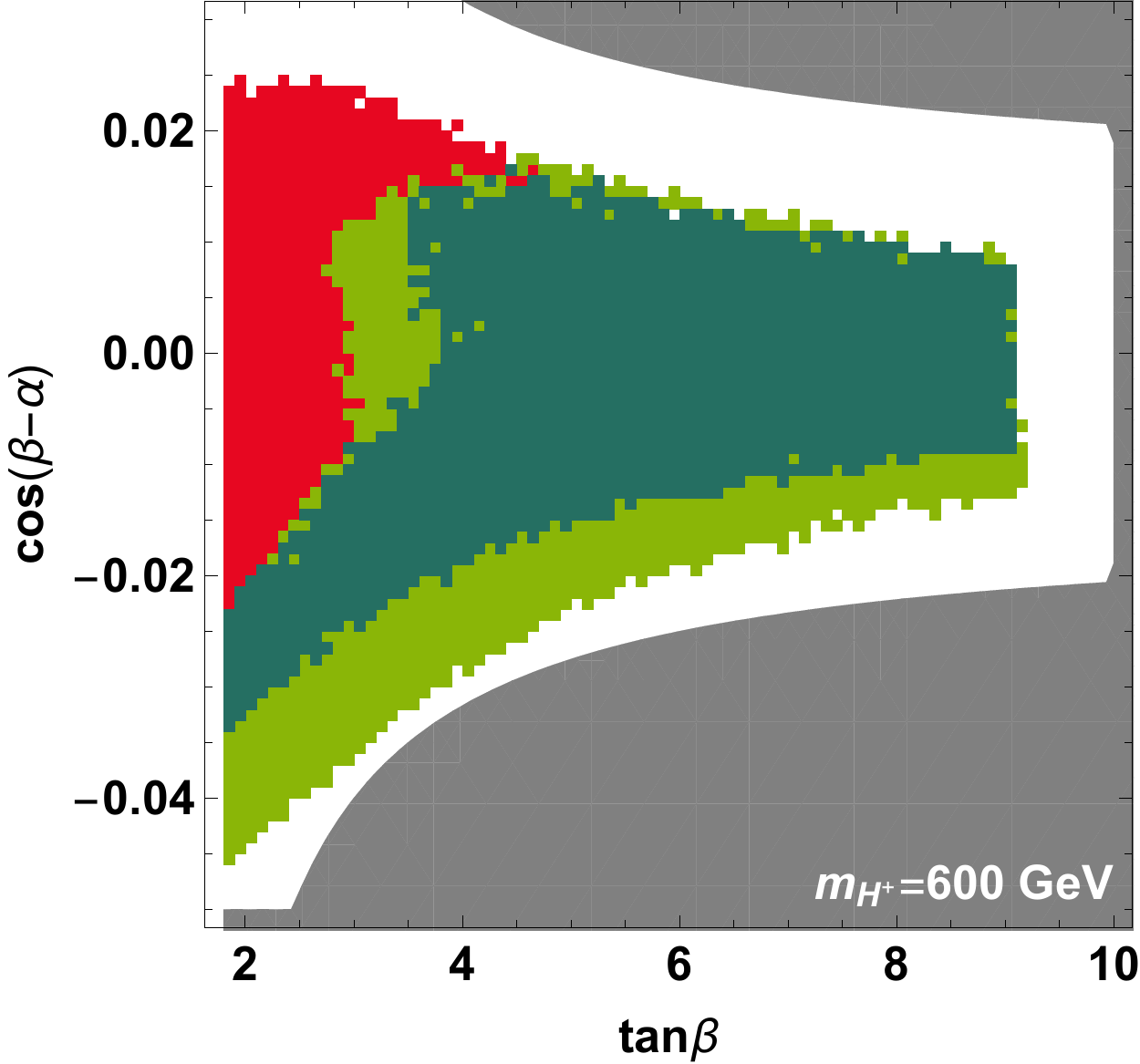}
\end{minipage}\hspace{1.5ex}
\begin{minipage}{0.3\linewidth}
 \includegraphics[width=\linewidth]{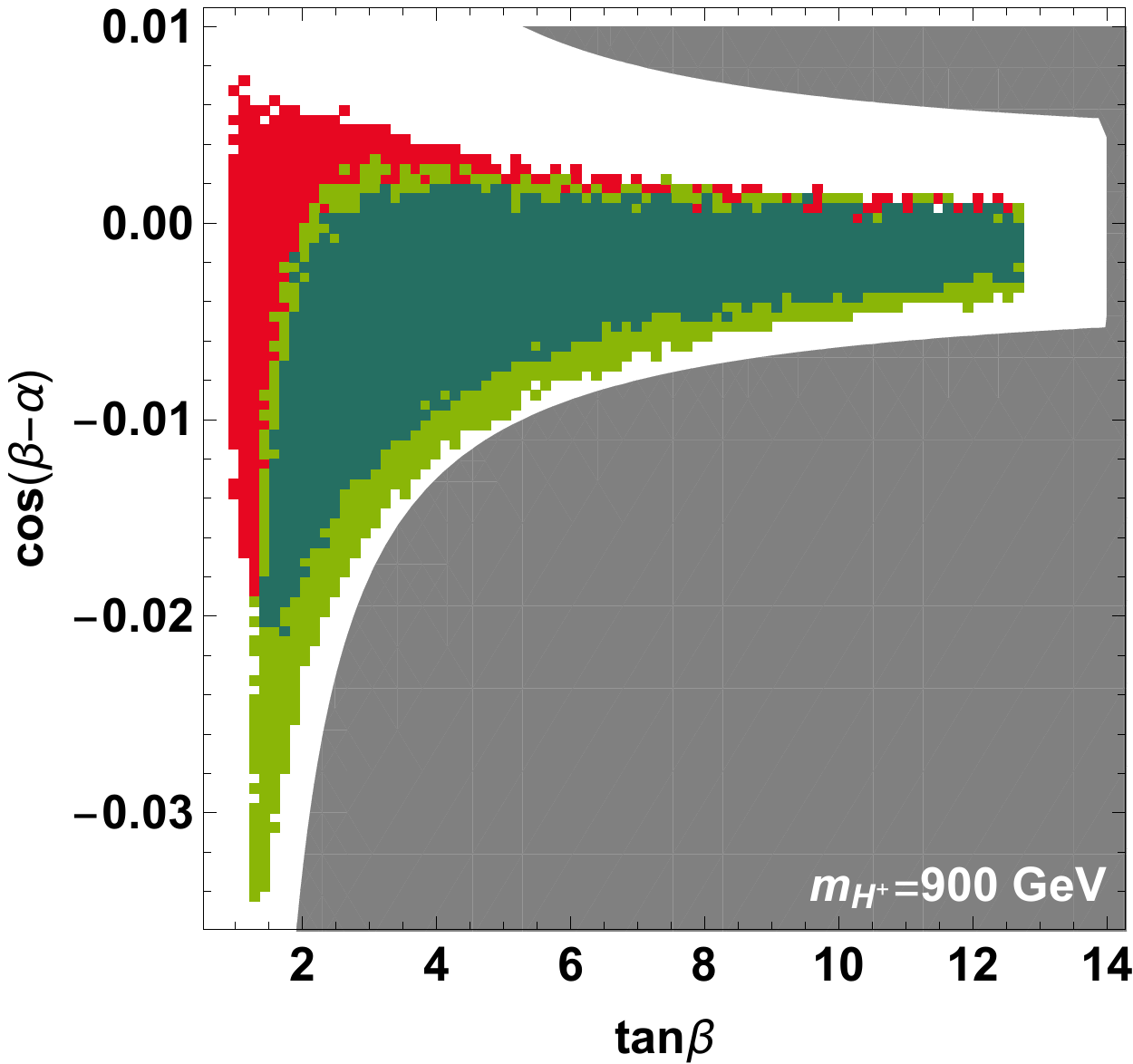}
\end{minipage}\hspace{1.5ex}
\begin{minipage}{0.3\linewidth}
 \includegraphics[width=\linewidth]{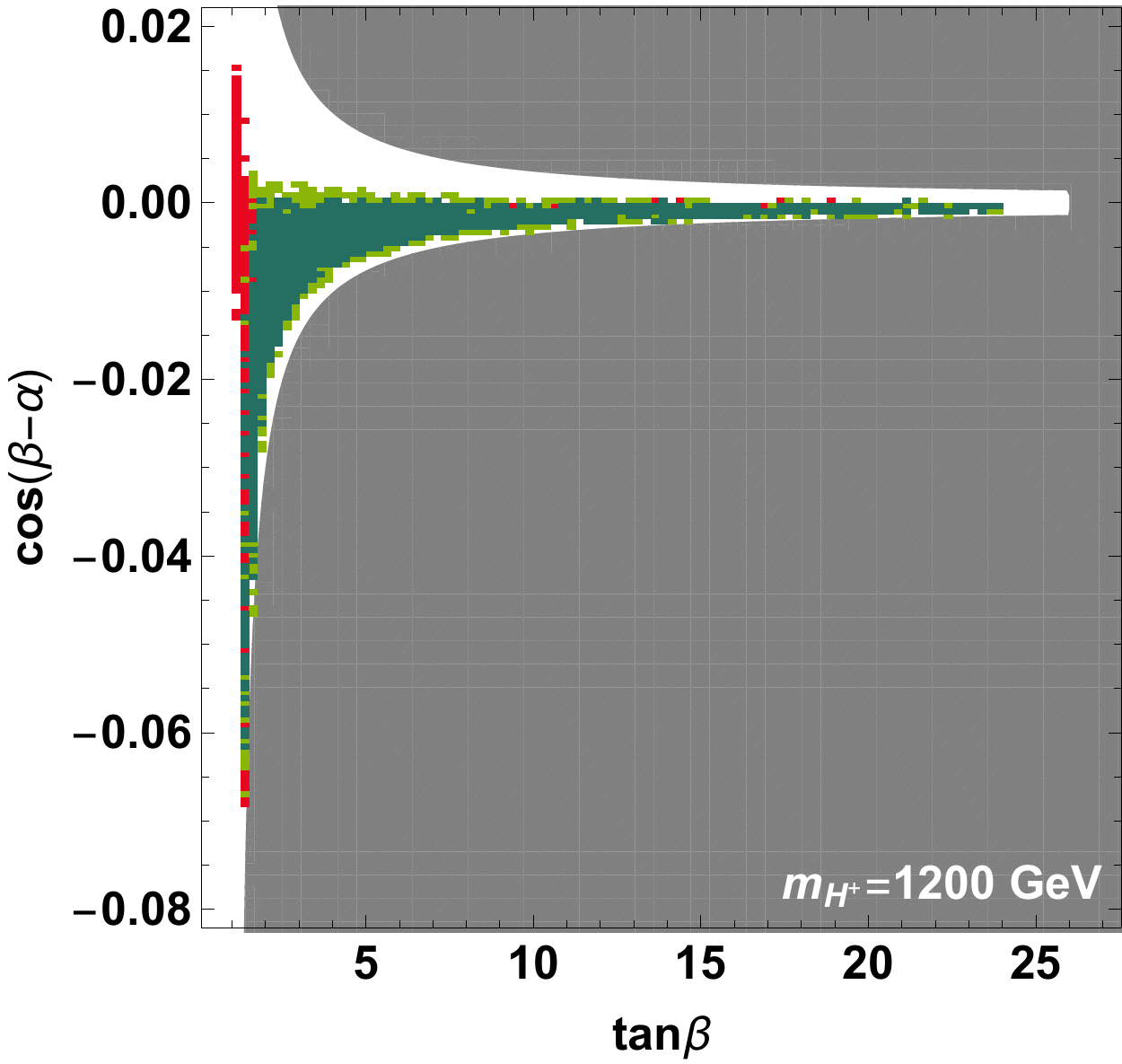}
\end{minipage}
\end{center}
\begin{center}
 \begin{minipage}{0.3\linewidth}
 \includegraphics[width=\linewidth]{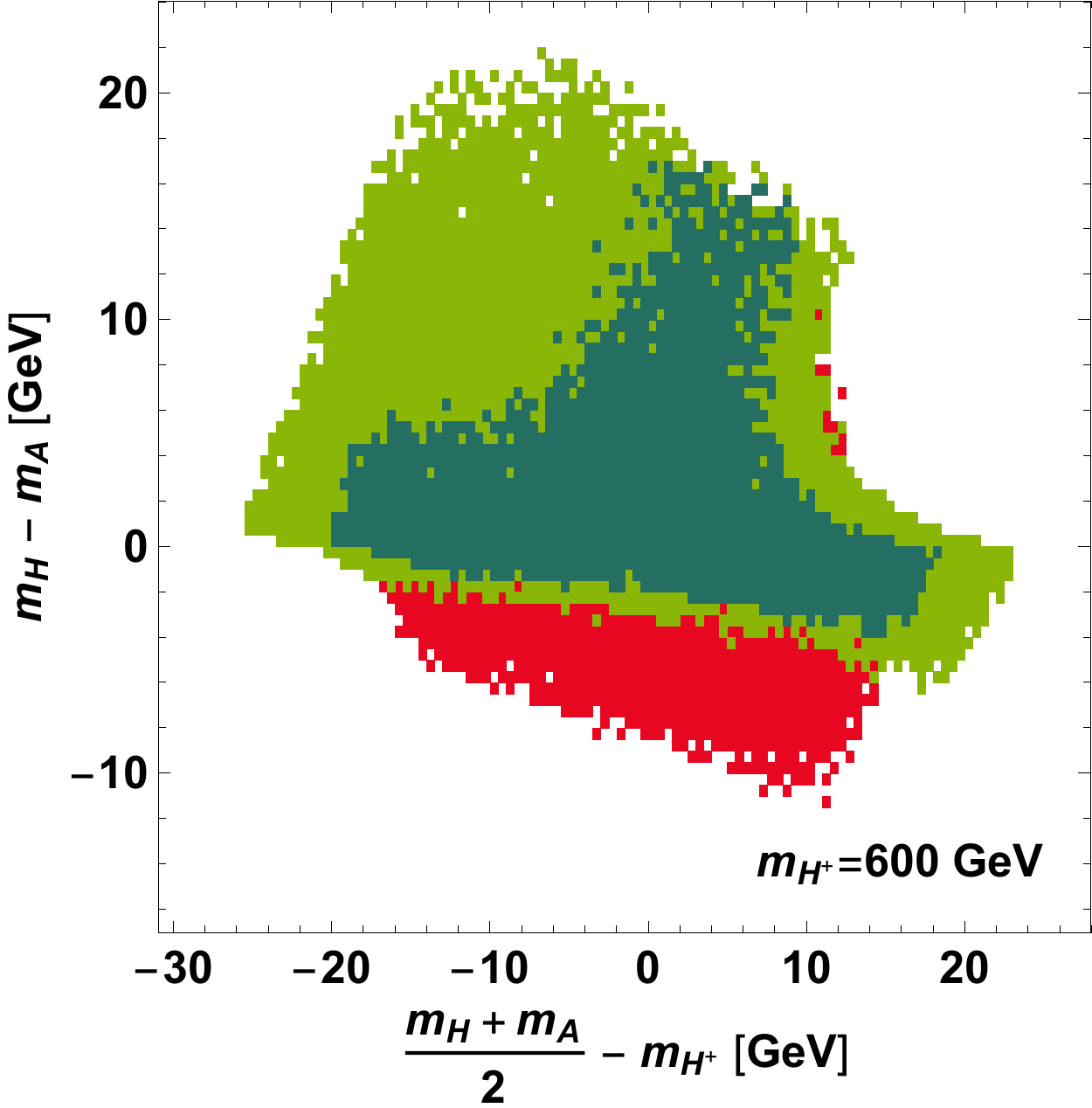}
\end{minipage}\hspace{1.5ex}
\begin{minipage}{0.3\linewidth}
 \includegraphics[width=\linewidth]{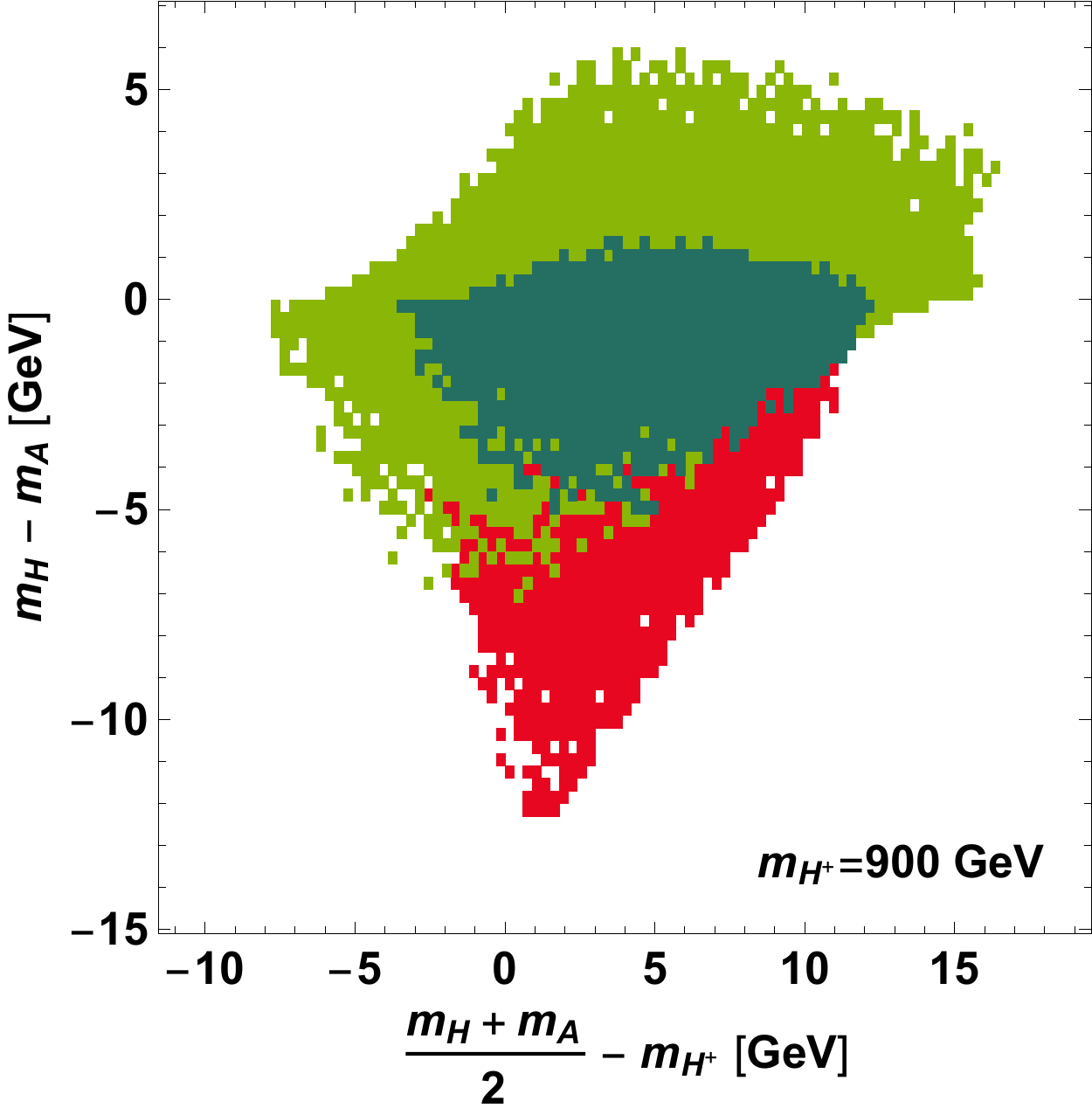}
\end{minipage}\hspace{1.5ex}
\begin{minipage}{0.3\linewidth}
 \includegraphics[width=\linewidth]{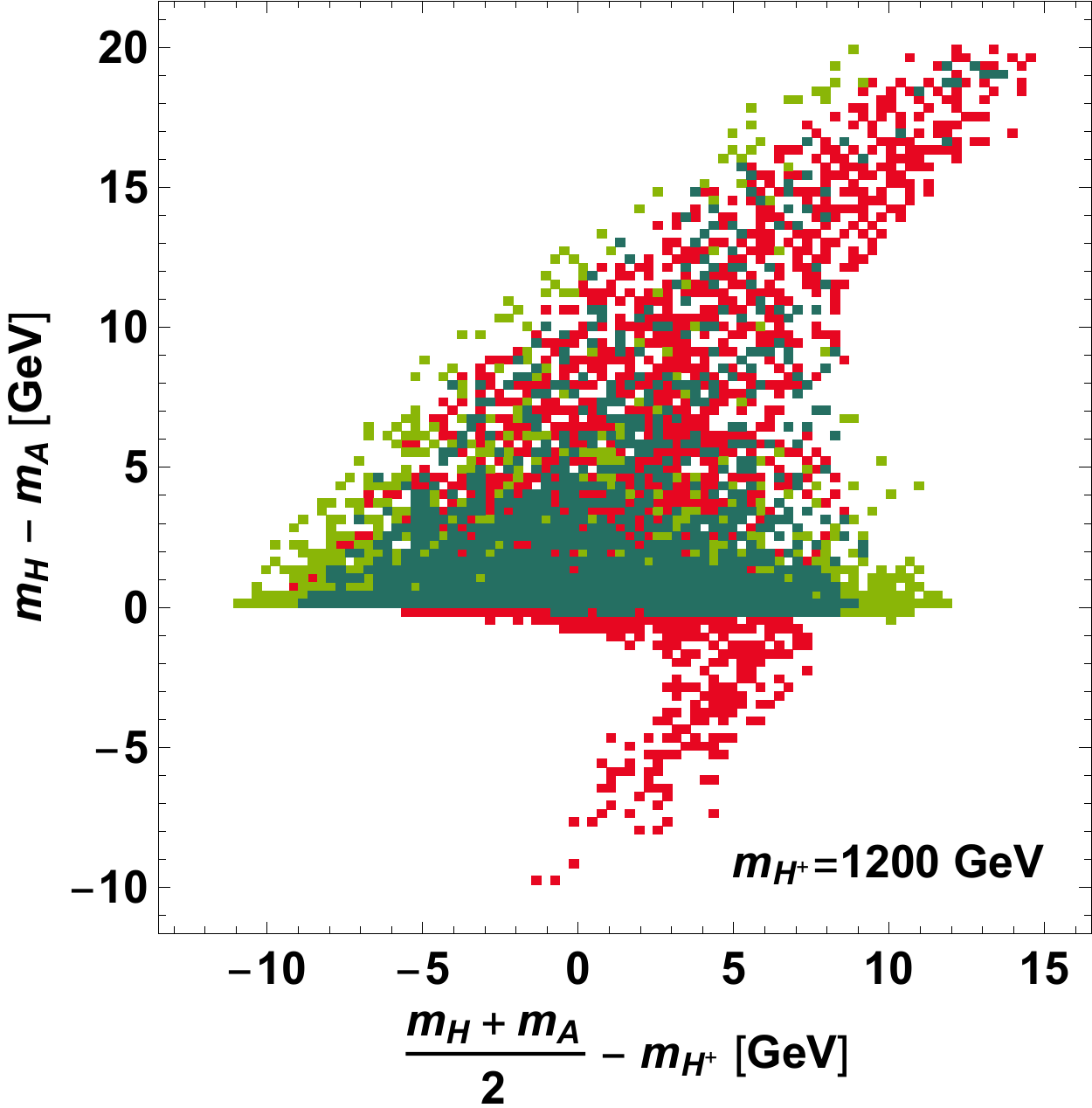}
\end{minipage}
\end{center}
\caption{The detail views of the parameter regions inside the white
dashed lines in Fig.~\ref{fig_scat1}. The masked region is shaded with
gray. All the points satisfy the low energy constraints and remain
perturbative until the Planck scale. The red points satisfy the
bounded-from-below constraint. The lighter green points satisfy the
vacuum stability constraint. The darker green points satisfy both of
them.} \label{fig_scat2}
\end{figure}

Let us discuss the implication on the Higgs couplings. At the tree level,
the SM-value normalized couplings of the $125~{\rm GeV}$ Higgs boson are
given by
\begin{align}
 g_{hUU}&=\sin(\beta-\alpha)+\cot\beta\cos(\beta-\alpha),\\
 g_{hDD}&=\sin(\beta-\alpha)-\tan\beta\cos(\beta-\alpha),\\
 g_{hLL}&=\sin(\beta-\alpha)-\tan\beta\cos(\beta-\alpha),\\
 g_{hVV}&=\sin(\beta-\alpha),
\end{align}
where $U,D,L$, and $V$ represent the up-type quarks, the down-type
quarks, the leptons, and the gauge bosons, respectively. Since
$|\cos(\beta-\alpha)|\lesssim0.06$ for all the slices, we have
$0.9982\lesssim g_{hVV}\leq1$, which is not possible to be distinguished
from unity even with HL-LHC plus $1~{\rm TeV}$ ILC
\cite{Fujii:2019zll}. It also means that the model cannot be valid up
to the Planck scale if we observe larger deviations of $g_{hVV}$
couplings. On the other hand, the other couplings can deviate by more
than $1\%$ because of the second term of the above equations.

\begin{figure}[t]
 \begin{center}
  \begin{minipage}{0.3\linewidth}
   \includegraphics[width=\linewidth]{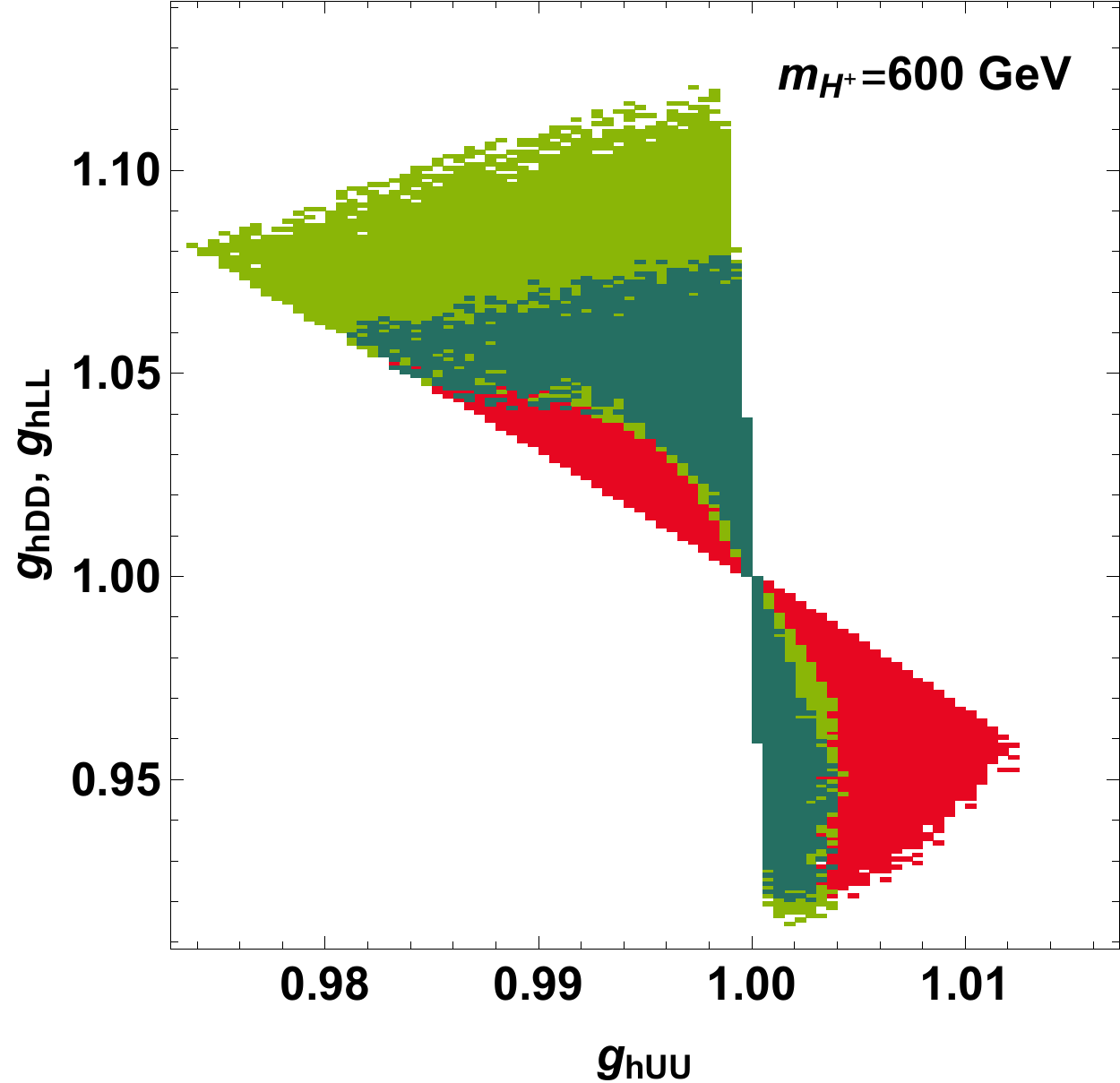}
  \end{minipage}\hspace{1.5ex}
  \begin{minipage}{0.3\linewidth}
   \includegraphics[width=\linewidth]{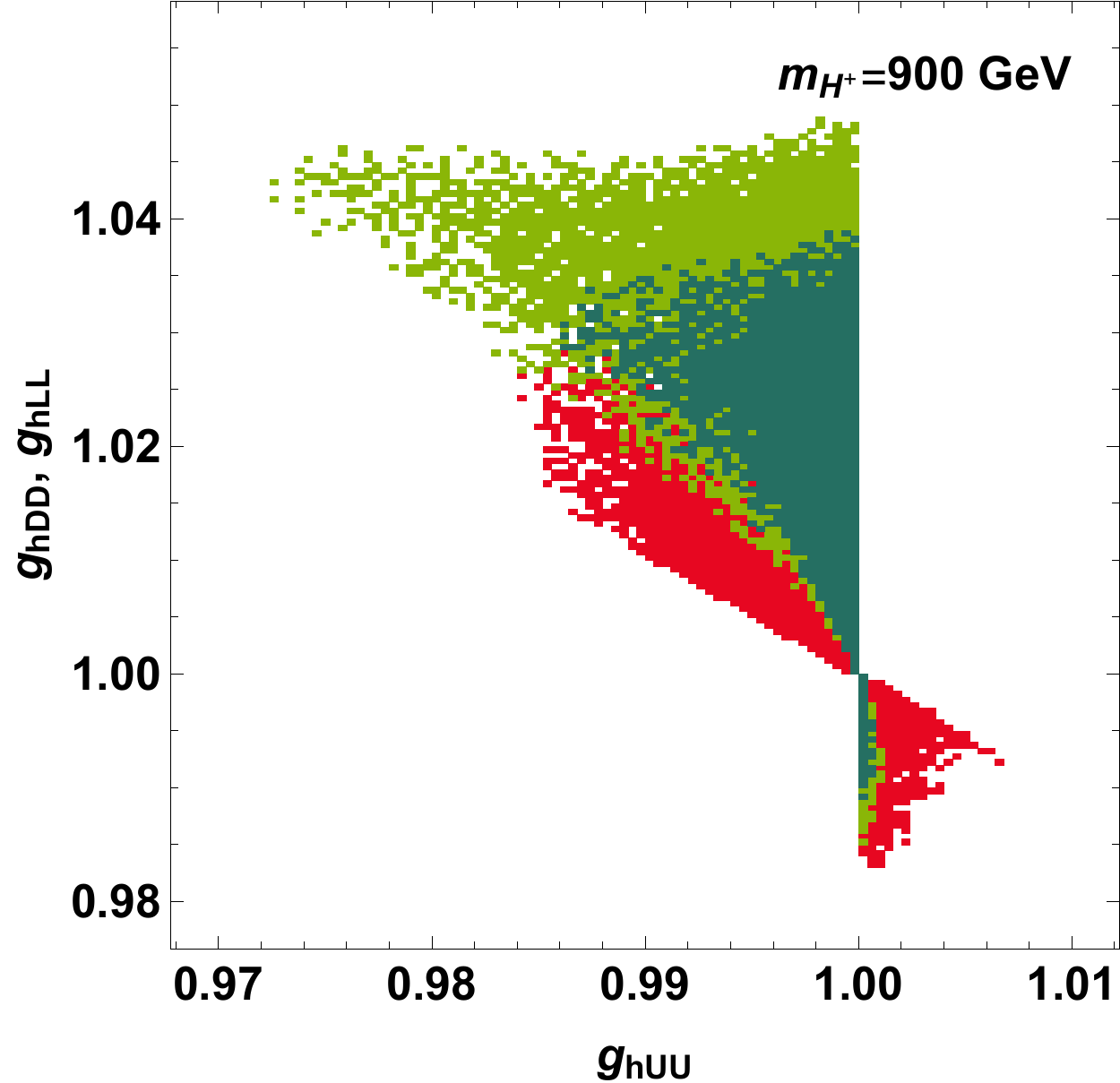}
  \end{minipage}\hspace{1.5ex}
  \begin{minipage}{0.3\linewidth}
   \includegraphics[width=\linewidth]{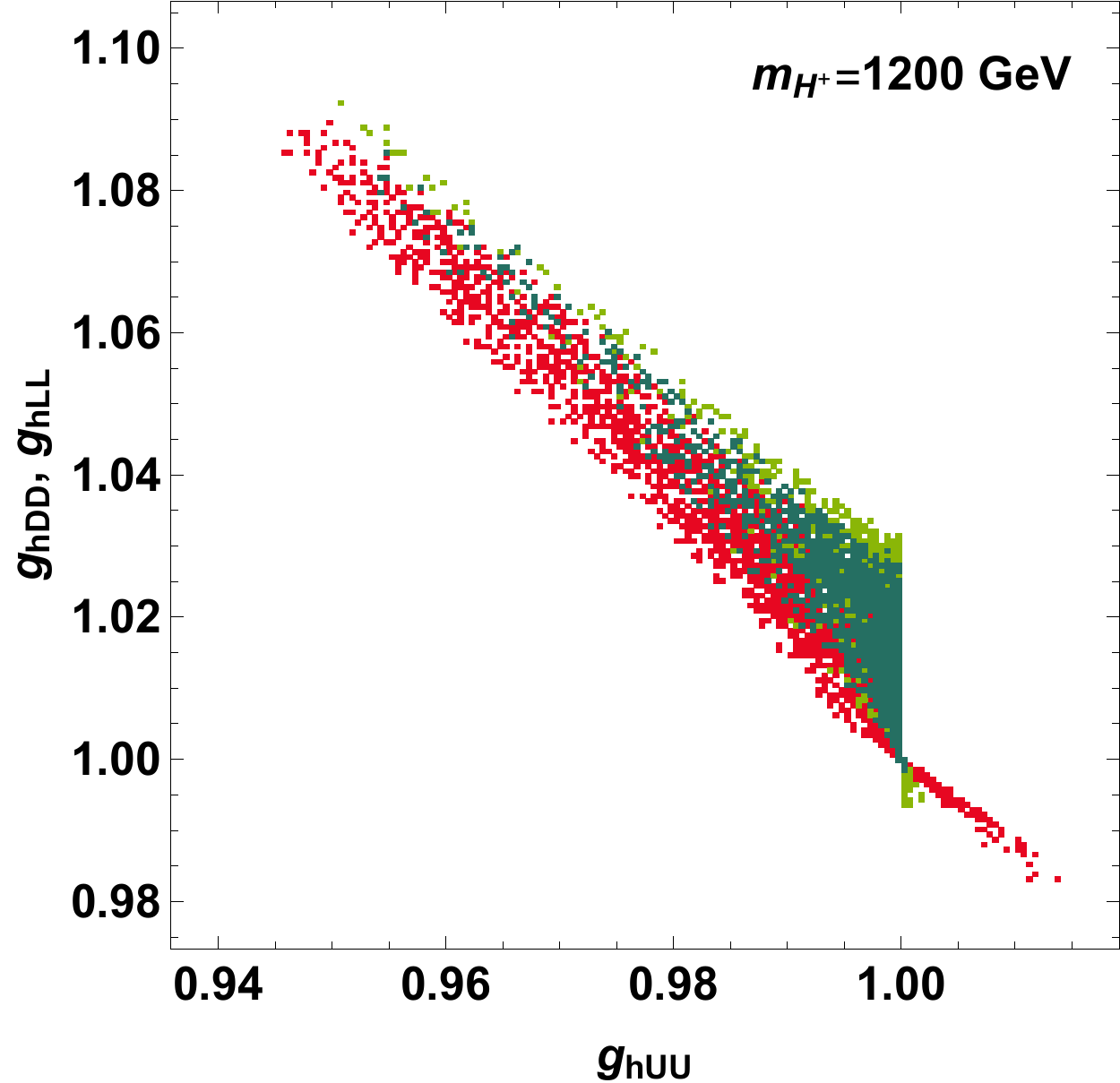}
  \end{minipage}
 \end{center}
\caption{The SM-normalized couplings of the $125~{\rm GeV}$ Higgs
boson. The colors are the same as in Fig.~\ref{fig_scat2}.}
\label{fig_hcoup}
\end{figure}

In Fig.~\ref{fig_hcoup}, we cast the data points in Fig.~\ref{fig_scat2}
into the $g_{hUU}$ vs $g_{hDD}=g_{hLL}$ plane. The colors are the same
as in Fig.~\ref{fig_scat2}. As we can see, $g_{hUU}$ can be reduced by
about $2\%-5\%$ for each slice, but cannot be enhanced so much. On
the other hand, $g_{hDD}$ and $g_{hLL}$ can deviate by about $5\%-12\%$
for each slice, and tend to be enhanced.

The current constraints on these couplings are given by \cite{Aad:2019mbh}
\begin{align}
 g_{hZZ}&=1.10\pm0.08,\\
 g_{hWW}&=1.05\pm0.08,\\
 g_{hbb}&=1.06^{+0.19}_{-0.18},\\
 g_{htt}&=1.02^{+0.11}_{-0.10},\\
 g_{h\tau\tau}&=1.07\pm0.15,
\end{align}
with the assumption that there is no new particles in loops and
decays. Thus, they have already started to touch the parameter space.
Future measurements of the Higgs couplings by, for example, the
combination of HL-LHC and ILC will reach the precision of a few percent
level \cite{Fujii:2019zll} and will possibly find deviations from the SM
values.

Let us discuss the dependence on the IR cut-off and the UV cut-off,
which are introduced in Eqs.~\eqref{eq_cutoff1} and
\eqref{eq_cutoff2}. Since the beta functions for $\lambda_1$ and
$\lambda_2$ become generically positive at high energy, a factor change
of the UV cut-off rarely affects the vacuum decay rate, which we have
checked numerically as well. As for the IR cut-off, we have checked that
Figs.~\ref{fig_scat2} and \ref{fig_hcoup} are not affected even if we
use $1~{\rm TeV}$ for the IR cut-off, instead of $10~{\rm TeV}$.

Finally, we comment on the effect of $\mathcal A$, which we have
calculated precisely. Although the vacuum decay rates are enhanced
compared with the tree level ones around $\gamma\sim H_0^4$, we find
that the effect is not large enough to change Fig.~\ref{fig_hcoup}. It
is because of the strong dependence of the vacuum decay rates on the
Higgs quartic couplings. However, if we find the additional Higgs bosons
in future, the vacuum decay rate can be determined precisely from the
measurements of the mass differences and the couplings of the Higgs
bosons, which will give an important implication on the scenario.
\section{Summary}\label{sec_summary}
In this paper, we analyzed the high scale validity of the DFSZ axion
model, namely the HE perturbative unitarity and the vacuum
stability. The model has been widely studied since it can explain the
strong CP problem and dark matter elegantly. Once we admit a mechanism
that forces classical scale invariance at the Planck scale, the Higgs
mass terms of the appropriate size can be generated through the
technically natural parameters and may be valid up to the Planck scale.
In addition, the model can be extended without affecting the Higgs
sector to explain the neutrino masses, the baryon asymmetry of the
Universe and inflation. Thus, the high scale behavior of the Higgs
sector is worth discussing.

We utilized the state-of-the-art method to calculate the vacuum decay
rate precisely. We extended the results of
\cite{Andreassen:2017rzq,Chigusa:2017dux,Chigusa:2018uuj} to accommodate
bounces that are composed of more than one fields. Then, we showed that
$\mathcal A$ can enhance the vacuum decay rates at most by $10^{10}$,
which can become comparable with the uncertainties from those of the top
mass and the strong coupling constant.

We performed the parameter scan and found the parameter space that
satisfies the constraints from flavor observables, LE/HE perturbative
unitarity, oblique parameters, collider searches, and vacuum
stability. Due to the complementarity of the HE perturbative unitarity
and the vacuum stability, the allowed parameter space becomes very
small. We observe that it still accommodates at most $12\%$ enhancement
of the $hDD$ and $hLL$ couplings, and at most $5\%$ suppression of the
$hUU$ couplings. These are around the current experimental constraints
and will be searched at future experiments such as HL-LHC and ILC. On
the other hand, the deviation of the $hVV$ couplings are found to be
smaller than $0.2\%$ and the scenario may be excluded if we observe
large deviations of these couplings.

\begin{acknowledgements}
Y.S.~is supported by Grant-in-Aid for Scientific research from the
Ministry of Education, Science, Sports, and Culture (MEXT), Japan,
No.~16H06492. S.O.~and D.-s.T.~are supported by the mathematical and
theoretical physics unit (Hikami unit) of the Okinawa Institute of
Science and Technology Graduate University. S.O.~and D.-s.T.~are also
supported by Japan Society for the Promotion of Science (JSPS),
Grant-in-Aid for Scientific Research (C), Grant Number JP18K03661. The
authors thank the Yukawa Institute for Theoretical Physics at Kyoto
University, where this work was initiated during the YITP-W-18-05 on
"Progress in Particle Physics 2018 (PPP2018)".
\end{acknowledgements}

\appendix
\section{Flavor}\label{apx_flavor}
In this appendix, we follow \cite{Enomoto:2015wbn} and obtain the flavor
constraints with the current experimental values.
\subsection{CKM Matrix Elements}
We first determine the CKM matrix elements by using observables that are
insensitive to the additional Higgs bosons.

We use the Wolfenstein parametrization defined as
\begin{equation}
 V_{\rm CKM}=
\begin{pmatrix}
 1-\frac{\lambda_{\rm CKM}^2}{2}&\lambda_{\rm CKM}&A_{\rm CKM}\lambda_{\rm CKM}^3(\rho_{\rm CKM}-i\eta_{\rm CKM})\\
 -\lambda_{\rm CKM}&1-\frac{\lambda_{\rm CKM}^2}{2} & A_{\rm CKM}\lambda_{\rm CKM}^2\\
 A_{\rm CKM}\lambda_{\rm CKM}^3(1-\rho_{\rm CKM}-i\eta_{\rm CKM})& -A_{\rm CKM}\lambda_{\rm CKM}^2&1 
\end{pmatrix},
\end{equation}
where we neglect $\mathcal O(\lambda_{\rm CKM}^4)$. We determine
$\lambda_{\rm CKM}$ by using the super allowed nuclear beta decays,
$|V_{ud}|=0.97420\pm0.00021$ \cite{Hardy:2016vhg}, and the $K\to e\nu$
decay, ${\rm BR}(K\to e\nu)=(1.582\pm0.007)\times10^{-5}$
\cite{Tanabashi:2018oca}. Combining these two, we obtain
\begin{align}
 \lambda_{\rm CKM}&=0.2244\pm0.0005.
\end{align}
Here, we have used experimental values for the $K$ meson in Table
\ref{tbl_param_meson}.

Next, we determine $A_{\rm CKM}$ from $|V_{cb}|$ assuming that the
corrections from the charged Higgs boson are small, which is justified
for $m_{H^+}>150~{\rm GeV}$ and $\tan\beta<100$
\cite{Enomoto:2015wbn}. We use $|V_{cb}|=(39.25\pm0.56)\times10^{-3}$
\cite{Amhis:2019ckw} and get
\begin{equation}
 A_{\rm CKM}=0.779\pm0.012.
\end{equation}

Finally, we determine $\rho_{\rm CKM}$ and $\eta_{\rm CKM}$ from the
unitary triangle. From $\phi_1=(22.2\pm0.7)^\circ$, $\phi_2=(84.9\pm5)^\circ$,
and $\phi_3=(71.1\pm5)^\circ$ \cite{Amhis:2019ckw}, we get
\begin{align}
 \bar\rho_{\rm CKM}&=0.117\pm0.020,\\
 \bar\eta_{\rm CKM}&=0.361\pm0.012,
\end{align}
where
\begin{equation}
 \rho_{\rm CKM}+i\eta_{\rm CKM}=\frac{\bar\rho_{\rm CKM}+i\bar\eta_{\rm CKM}}{1-A_{\rm CKM}^2\lambda_{\rm CKM}^4(\bar\rho_{\rm CKM}+i\bar\eta_{\rm CKM})}\sqrt{\frac{1-A_{\rm CKM}^2\lambda_{\rm CKM}^4}{1-\lambda_{\rm CKM}^2}}.
\end{equation}

\begin{table}[t]
 \begin{tabular}{cc}
\hline\hline
  \multicolumn{2}{c}{Mesons}\\
\hline\hline
  Input&Value \\
\hline
  $m_{K^\pm}$&$493.677~{\rm MeV}$ \cite{Tanabashi:2018oca}\\
  $m_{B^\pm}$&$5.27933~{\rm GeV}$ \cite{Tanabashi:2018oca}\\
  $m_{B_s}$&$5.36688~{\rm GeV}$ \cite{Tanabashi:2018oca}\\
  $\tau_{K^\pm}$&$12.38\pm0.02~{\rm ns}$ \cite{Tanabashi:2018oca}\\
  $\tau_{B^\pm}$&$1.638\pm0.004~{\rm ps}$ \cite{Tanabashi:2018oca}\\
  $\tau_{B_s}^H$&$1.619\pm0.009~{\rm ps}$ \cite{Tanabashi:2018oca}\\
  $\tau_{B_s}^L$&$1.414\pm0.006~{\rm ps}$ \cite{Tanabashi:2018oca}\\
\hline\hline\hline
 \end{tabular}
 \begin{tabular}{cc}
\hline\hline
  \multicolumn{2}{c}{Theoretical Inputs}\\
\hline\hline
  Input&Value \\
\hline
  ${\rm BR}(b\to s\gamma)^{\rm SM}_{E_\gamma>1.6{\rm GeV}}$&$(3.36\pm0.23)\times10^{-4}$ \cite{Misiak:2015xwa,Czakon:2015exa}\\
  $f_K$&$155.7\pm0.3~{\rm MeV}$ \cite{Aoki:2019cca}\\
  $f_B$&$190.0\pm1.3~{\rm MeV}$ \cite{Aoki:2019cca}\\
  $f_{B_s}$&$230.3\pm1.3~{\rm MeV}$ \cite{Aoki:2019cca}\\
  $f_{B_s}^2B^{(s)}_2(m_b)$&$0.0421\pm0.0028~{\rm GeV}^2$ \cite{Bazavov:2016nty}\\
  $f_{B_s}^2B^{(s)}_3(m_b)$&$0.0576\pm0.0078~{\rm GeV}^2$ \cite{Bazavov:2016nty}\\
  $\hat B_{B_s}$&$1.35\pm0.06$ \cite{Aoki:2019cca}\\
\hline\hline\hline
 \end{tabular}
\caption{Experimental and theoretical parameters for mesons. The
uncertainties are used for the evaluation of theoretical uncertainties
in the flavor analysis.}  \label{tbl_param_meson}
\end{table}

\begin{table}[t]
 \begin{tabular}{cc}
\hline\hline
  \multicolumn{2}{c}{EW Parameters}\\
\hline\hline
  Input&Value\\
\hline
  $m_h$&$125.1\pm0.14~{\rm GeV}$ \cite{Tanabashi:2018oca}\\
  $m_W$&$80.379~{\rm GeV}$ \cite{Tanabashi:2018oca}\\
  $m_Z$&$91.1876~{\rm GeV}$ \cite{Tanabashi:2018oca}\\
  $\alpha_s(m_Z)$&$0.1181$ \cite{Tanabashi:2018oca}\\
  $G_F$&$1.1663787\times10^{-5}~{\rm GeV}^{-2}$ \cite{Tanabashi:2018oca}\\
  $\sin^2\theta_W(m_Z)$&$0.23122$ \cite{Tanabashi:2018oca}\\
\hline\hline\hline
 \end{tabular}
 \begin{tabular}{cc}
\hline\hline
  \multicolumn{2}{c}{Fermion Masses}\\
\hline\hline
  Input&Value\\
\hline
  $M_t$&$173.1\pm0.9~{\rm GeV}$ \cite{Tanabashi:2018oca}\\
  $m_b(m_b)$&$4.198\pm0.012~{\rm GeV}$ \cite{Aoki:2019cca}\\
  $m_s(2~{\rm GeV})$&$93.44\pm0.68~{\rm MeV}$ \cite{Aoki:2019cca}\\
  $m_u(2~{\rm GeV})$&$2.50\pm0.17~{\rm MeV}$ \cite{Aoki:2019cca}\\
  $m_\mu$&$105.6583745~{\rm MeV}$ \cite{Tanabashi:2018oca}\\
  $m_\tau$&$1.77686~{\rm GeV}$ \cite{Tanabashi:2018oca}\\
\hline\hline\hline
 \end{tabular}
\caption{Fundamental parameters of the SM. The uncertainties are used
for the evaluation of theoretical uncertainties in the flavor
analysis. The top mass is the on-shell mass and the other quark
masses are the $\overline{\rm MS}$ masses with the renormalization scale
shown in the parentheses.}  \label{tbl_param_sm}
\end{table}

\begin{table}[t]
 \begin{tabular}{cc}
\hline\hline
  \multicolumn{2}{c}{Experimental Results}\\
\hline\hline
  Observable&Value\\
\hline
  ${\rm BR}(b\to s\gamma)_{E_\gamma>1.6{\rm GeV}}$&$(3.32\pm0.15)\times10^{-4}$ \cite{Amhis:2019ckw}\\
  ${\rm BR}(B\to\tau\nu)$&$(1.06\pm0.19)\times10^{-4}$ \cite{Amhis:2019ckw}\\
  ${\rm BR}(B_s\to\mu\mu)$&$(3.1\pm0.6)\times10^{-9}$ \cite{Amhis:2019ckw}\\
  $\Delta M_{B_s}$&$17.757\pm0.021~{\rm ps}^{-1}$ \cite{Amhis:2019ckw}\\
\hline\hline\hline
 \end{tabular}
\caption{Experimental results for flavor observables.}
\label{tbl_flavor}
\end{table}
\subsection{Flavor Constraints}
For the theoretical evaluation of the flavor observables, we use the
formulas given in \cite{Enomoto:2015wbn}. In the calculation, we utilize
{\tt RunDec} \cite{Chetyrkin:2000yt,Herren:2017osy} to get the running
masses of quarks.

Let us clarify the statistical method that we adopt. For observable $X$
that depends on known parameters $\{x_i\pm\delta x_i\}$ and model
parameters $\{y_i\}$, we define
\begin{equation}
  \chi^2(\{y_i\})=\frac{(X(\{y_i\})-X_{\rm exp})^2}{\delta X_{\rm th}^2(\{y_i\})+\delta X_{\rm exp}^2},
\end{equation}
where $X_{\rm exp}\pm\delta X_{\rm exp}$ is the experimental result,
$X(\{y_i\})$ is the theoretical result for inputs $\{x_i\}$ and
$\{y_i\}$, and
\begin{equation}
 \delta X_{\rm th}^2(\{y_i\})=\sum_k \left[X(\{y_i\})|_{x_k\to x_k+\delta x_k/2}-X(\{y_i\})|_{x_k\to x_k-\delta x_k/2}\right]^2.
\end{equation}
Then, we use the offset corrected $\chi^2$ defined as
\begin{equation}
 \Delta \chi^2(\{y_i\})=\chi^2(\{y_i\})-\min_{\{\tilde y_i\}}\chi^2(\{\tilde y_i\}),
\end{equation}
and the $95\%$ CL exclusion limit is given by $\Delta
\chi^2(\{y_i\})\lesssim3.84$. Here, the minimum value is searched over
the parameter space of Fig.~\ref{fig_flavor} and the SM limit.  The known
parameters $\{x_i\pm\delta x_i\}$ are summarized in Tables
\ref{tbl_param_meson} and \ref{tbl_param_sm}, where we ignore
uncertainties of $\mathcal O(0.1\%)$\footnote{We have also ignored the
uncertainty of the strong coupling constant since its effect is
suppressed compared with other uncertainties.}.

The result is shown in Fig.~\ref{fig_flavor}.

\section{Proof of Straight Bounce}\label{apx_straight}
In this appendix, we show that $\Omega$ and $\Theta$ of the relevant
solution do not depend on $r$. In terms of $\phi$, $\Omega$ and
$\Theta$, the Euclidean action is expressed as
\begin{align}
 S_E[\phi,\Omega,\Theta]&=\mathcal K[\phi,\Omega,\Theta]+\mathcal V[\phi,\Omega,\Theta],\\
 \mathcal K[\phi,\Omega,\Theta]&=2\pi^2\int drr^3\left[\frac{1}{2}\phi'^2+\frac{1}{2}\phi^2\left(\Omega'^2+\Theta'^2\sin^2\Omega\right)\right],\\
 \mathcal V[\phi,\Omega,\Theta]&=2\pi^2\int drr^3\frac{\lambda_\phi}{4}\phi^4,
\end{align}
where $\Omega'$, $\Theta'$ and $\phi'$ are derivatives with respect to
$r$. As introduced in \cite{Coleman:1977th}, we can obtain the bounce by
minimizing $\mathcal K[\tilde\phi,\tilde\Omega,\tilde\Theta]$ with the
constraint given by
\begin{align}
 \mathcal V[\tilde\phi,\tilde\Omega,\tilde\Theta]&=({\rm const.})<0.
\end{align}
After the minimization, the bounce solution is obtained as
\begin{align}
 \phi(r)=\tilde\phi(\sigma r),~
 \Omega(r)=\tilde\Omega(\sigma r),~
 \Theta(r)=\tilde\Theta(\sigma r),
\end{align}
where
\begin{equation}
 \sigma=\sqrt{-\frac{2\mathcal V[\tilde\phi,\tilde\Omega,\tilde\Theta]}{\mathcal K[\tilde\phi,\tilde\Omega,\tilde\Theta]}}.
\end{equation}
Its Euclidean action is given by
\begin{equation}
 S_E=\frac{\mathcal K[\tilde\phi,\tilde\Omega,\tilde\Theta]}{2\sigma^2}.\label{eq_reduced_action}
\end{equation}

Let us assume that there exists a minimum, $\mathcal
K[\tilde\phi_A,\tilde\Omega_A,\tilde\Theta_A]$, where $\tilde\Omega_A$
or $\tilde\Theta_A$ is not constant. Since $\lambda_\phi(\Omega,\Theta)$
is a continuous function, there exist constant $\Omega_B$ and $\Theta_B$
satisfying
\begin{equation}
 \mathcal V[\tilde\phi_A,\tilde\Omega_A,\tilde\Theta_A]=\mathcal V[\tilde\phi_A,\Omega_B,\Theta_B].
\end{equation}
Then, we have
\begin{equation}
 \mathcal K[\tilde \phi_A,\tilde\Omega_A,\tilde\Theta_A]-\mathcal K[\tilde\phi_A,\Omega_B,\Theta_B]=2\pi^2\int drr^3\frac{1}{2}\tilde\phi_A^2\left(\tilde\Omega'^2_A+\tilde\Theta'^2_A\sin^2\tilde\Omega_A\right)\geq0.
\end{equation}
The equality holds only when $\tilde\Omega'(r)=\tilde\Theta'(r)=0$ for
any $r$. Notice that when $\sin\tilde\Omega_A=0$, the field space is not
parameterized by $\tilde\Theta_A$. Then, from
Eq.~\eqref{eq_reduced_action}, there exists a bounce with smaller action
if $\tilde\Omega$ or $\tilde\Theta$ is not constant.  Thus, the bounce
with minimum action can only be realized with constant $\Omega$ and
$\Theta$\footnote{Only the bounce with minimum action is relevant for
the vacuum decay since the contributions from the others are
exponentially suppressed.}.

\section{One-loop Corrections to a Vacuum Decay Rate}\label{apx_oneloop}
From \cite{Chigusa:2018uuj}, the differential vacuum decay rate is
expressed as
\begin{align}
 \frac{d\gamma}{dR}&=\left.\frac{1}{R^5}\mathcal A'^{(h)}\mathcal A^{(\sigma)}\mathcal A^{(\psi)}\mathcal A^{(A_\mu,\varphi)}e^{-\mathcal B}\right|_{\mu\sim R^{-1}},
\end{align}
where 
\begin{align}
 \ln\mathcal A'^{(h)}&=\left.\left[\ln\mathcal A'^{(h)}\right]_{\overline{\rm MS}}\right|_{\lambda\to\lambda_\phi},\\
 \ln\mathcal A^{(\sigma)}&=\sum_i\left.n_i^{(\sigma)}\left[\ln\mathcal A^{(\sigma)}\right]_{\overline{\rm MS}}\right|_{\kappa\to\kappa_i,~\lambda\to\lambda_\phi},\\
 \ln\mathcal A^{(\psi)}&=\sum_i\left.n_i^{(\psi)}\left[\ln\mathcal A^{(\psi)}\right]_{\overline{\rm MS}}\right|_{y\to y_i,~\lambda\to\lambda_\phi},\\
 \ln\mathcal A^{(A_\mu,\varphi)}&=\ln\mathcal V_G+\sum_i\left.n_i^{(A_\mu,\varphi)}\left[\ln\mathcal A'^{(A_\mu,\varphi)}\right]_{\overline{\rm MS}}\right|_{g^2\to \tilde g_i^2,~\lambda\to\lambda_\phi}.
\end{align}
Here, $[\ln\mathcal A^{(X)}]_{\overline{\rm MS}}$'s are defined in
\cite{Chigusa:2018uuj}. The degrees of freedom, $n_i^{(X)}$, and the
couplings, $\kappa_i$, $y_i$ and $g_i^2$, are summarized below for each
case. For case (c), the symmetry breaking pattern depends on the sign of
$\lambda_4$. Thus, we divide it into two cases; (c.1): $\lambda_4<0$ and
(c.2): $\lambda_4>0$.
\begin{description}
 \item[case (a)] $\lambda_\phi=\frac{1}{2}\lambda_1$

The scalar contributions:
\begin{align}
 n_1^{(\sigma)}=2,&~\kappa_1=\frac{\lambda_3}{2},\\
 n_2^{(\sigma)}=2,&~\kappa_2=\frac{\lambda_3+\lambda_4}{2}.
\end{align}
The fermion contributions:
\begin{align}
 n_1^{(\psi)}=3,&~y_1=y_b,\\
 n_2^{(\psi)}=1,&~y_2=y_\tau.
\end{align}
The gauge boson contributions:
\begin{align}
 n_1^{(A_\mu,\varphi)}=2,&~\tilde g_1^2=\frac{g_2^2}{4},\\
 n_2^{(A_\mu,\varphi)}=1,&~\tilde g_2^2=\frac{g_Y^2+g_2^2}{4}.
\end{align}
 \item[case (b)] $\lambda_\phi=\frac{1}{2}\lambda_2$

The scalar contributions:
\begin{align}
 n_1^{(\sigma)}=2,&~\kappa_1=\frac{\lambda_3}{2},\\
 n_2^{(\sigma)}=2,&~\kappa_2=\frac{\lambda_3+\lambda_4}{2}.
\end{align}
The fermion contributions:
\begin{align}
 n_1^{(\psi)}=3,&~y_1=y_t.
\end{align}
The gauge boson contributions:
\begin{align}
 n_1^{(A_\mu,\varphi)}=2,&~\tilde g_1^2=\frac{g_2^2}{4},\\
 n_2^{(A_\mu,\varphi)}=1,&~\tilde g_2^2=\frac{g_Y^2+g_2^2}{4}.
\end{align}
 \item[case (c.1)] $\lambda_\phi=\frac{1}{2}\frac{\lambda_1\lambda_2-(\lambda_3+\lambda_4)^2}{\lambda_1+\lambda_2-2(\lambda_3+\lambda_4)}$

The scalar contributions:
\begin{align}
 n_1^{(\sigma)}=2,&~\kappa_1=\lambda_\phi+\frac{|\lambda_4|}{2},\\
 n_2^{(\sigma)}=1,&~\kappa_2=\lambda_\phi+\frac{(\lambda_3+\lambda_4-\lambda_1)(\lambda_3+\lambda_4-\lambda_2)}{\lambda_1+\lambda_2-2(\lambda_3+\lambda_4)},\\
 n_3^{(\sigma)}=1,&~\kappa_3=\lambda_\phi.
\end{align}
The fermion contributions:
\begin{align}
 n_1^{(\psi)}=3,&~y_1=y_t\sqrt{\frac{\lambda_1-\lambda_3-\lambda_4}{\lambda_1+\lambda_2-2(\lambda_3+\lambda_4)}},\\
 n_2^{(\psi)}=3,&~y_2=y_b\sqrt{\frac{\lambda_2-\lambda_3-\lambda_4}{\lambda_1+\lambda_2-2(\lambda_3+\lambda_4)}},\\
 n_3^{(\psi)}=1,&~y_3=y_\tau\sqrt{\frac{\lambda_2-\lambda_3-\lambda_4}{\lambda_1+\lambda_2-2(\lambda_3+\lambda_4)}}.
\end{align}
The gauge boson contributions:
\begin{align}
 n_1^{(A_\mu,\varphi)}=2,&~\tilde g_1^2=\frac{g_2^2}{4},\\
 n_2^{(A_\mu,\varphi)}=1,&~\tilde g_2^2=\frac{g_Y^2+g_2^2}{4}.
\end{align}
 \item[case (c.2)] $\lambda_\phi=\frac{1}{2}\frac{\lambda_1\lambda_2-\lambda_3^2}{\lambda_1+\lambda_2-2\lambda_3}$

The scalar contributions:
\begin{align}
 n_1^{(\sigma)}=2,&~\kappa_1=\lambda_\phi+\frac{|\lambda_4|}{2},\\
 n_2^{(\sigma)}=1,&~\kappa_2=\lambda_\phi+\frac{(\lambda_3-\lambda_1)(\lambda_3-\lambda_2)}{\lambda_1+\lambda_2-2\lambda_3}.
\end{align}
The fermion contributions:
\begin{align}
 n_1^{(\psi)}=3,&~y_1=\sqrt{y_t^2\frac{\lambda_1-\lambda_3}{\lambda_1+\lambda_2-2\lambda_3}+y_b^2\frac{\lambda_2-\lambda_3}{\lambda_1+\lambda_2-2\lambda_3}},\\
 n_2^{(\psi)}=1,&~y_2=y_\tau\sqrt{\frac{\lambda_2-\lambda_3}{\lambda_1+\lambda_2-2\lambda_3}}.
\end{align}
The gauge boson contributions:
\begin{align}
 n_1^{(A_\mu,\varphi)}=2,&~\tilde g_1^2=\frac{g_2^2}{4},\\
 n_2^{(A_\mu,\varphi)}=1,&~\tilde g_2^2=\frac{g_Y^2+g_2^2+\sqrt{(g_Y^2-g_2^2)^2+4g_Y^2g_2^2\left(\frac{\lambda_1-\lambda_2}{\lambda_1+\lambda_2-2\lambda_3}\right)^2}}{8},\\
 n_3^{(A_\mu,\varphi)}=1,&~\tilde g_3^2=\frac{g_Y^2+g_2^2-\sqrt{(g_Y^2-g_2^2)^2+4g_Y^2g_2^2\left(\frac{\lambda_1-\lambda_2}{\lambda_1+\lambda_2-2\lambda_3}\right)^2}}{8}.
\end{align}
\end{description}
Here, $g_Y$ and $g_2$ are the gauge couplings for $U(1)_Y$ and
$SU(2)_L$, respectively. The group volume is $\mathcal V_G=2\pi^2$ for
cases (a), (b) and (c.1), and $\mathcal V_G=4\pi^3$ for case (c.2).

Notice that we can determine whether a solution, $\phi$, is a minimum or
not from the sign of $\kappa_i-\lambda_\phi$. Let $\chi_i$ be a scalar
orthogonal to $\phi$. Then, its potential can be written as
\begin{equation}
 V(\chi_i)=\frac{\lambda_\phi}{4}(\phi^2+\chi_i^2)^2+\frac{\delta}{2}\phi^2\chi_i^2+\dots,
\end{equation}
where $\delta$ breaks the rotational symmetry of $(\phi,\chi_i)$. Since
$\kappa_i$ can be read off from the mass term for $\chi_i$, we get
\begin{equation}
 \kappa_i=\lambda_\phi+\delta.
\end{equation}
Thus, for $\phi$ to be a minimum of the action, we need
$\kappa_i-\lambda_\phi>0$ for all $i$.

For case (c.1), we have $\kappa_3=\lambda_\phi$ and thus there appears a
zero mode, which is due to the spontaneous breaking of the PQ
symmetry. Its treatment is discussed in Appendix E of
\cite{Chigusa:2018uuj} and we replace
\begin{equation}
 \left.n_3^{(\sigma)}\left[\ln\mathcal A^{(\sigma)}\right]_{\overline{\rm MS}}\right|_{\kappa\to\kappa_3,\lambda\to\lambda_\phi}\to\mathcal V_\sigma\left.n_3^{(\sigma)}\left[\ln\mathcal A'^{(A_\mu,\varphi)}\right]_{\overline{\rm MS}}\right|_{g^2\to0,~\lambda\to\lambda_\phi},
\end{equation}
with $\mathcal V_\sigma=2\pi$.

\section{Matching Conditions}\label{apx_matching}
This appendix is devoted to the explanation of the one-loop matching
conditions for the dimensionless coupling constants. The matching scale
is taken to be $\mu_t=m_t$. For a detailed discussion of the
renormalization scheme, see \cite{Denner:1991kt,Altenkamp:2017ldc}. We
assume that $\tan\beta$ and $\cos(\beta-\alpha)$ are renormalized with
the $\msbar$ scheme. In the calculation of one-loop threshold
corrections, we utilize the public codes of {\tt SARAH}
\cite{Staub:2011dp,Staub:2015kfa}, {\tt FeynArts} \cite{Hahn:2000kx},
{\tt FeynCalc} \cite{Mertig:1990an,Shtabovenko:2016sxi}.
\subsection{Gauge Couplings}
We first evaluate the electric charge at $\mu_t$ as
\begin{equation}
 \frac{[e^{\rm SM(5)}(\mu_t)]^2}{4\pi}=\frac{\alpha^{\rm SM(5)}(m_Z)}{1-\frac{b^{\rm SM(5)}_e}{2\pi}\alpha^{\rm SM(5)}(m_Z)\ln\frac{\mu_t}{m_Z}},
\end{equation}
where \cite{Tanabashi:2018oca}
\begin{align}
 [\alpha^{{\rm SM}(5)}(m_Z)]^{-1}&=127.955\pm0.010,\\
 b^{\rm SM(5)}_e&=\frac{38}{9}.
\end{align}
It is then matched to the THDM electric charge as
\begin{equation}
 e(\mu_t)=\frac{e^{\rm SM(5)}(\mu_t)}{1-\Delta e},
\end{equation}
where
\begin{equation}
 \Delta e=-\frac{[e^{\rm SM(5)}(\mu_t)]^2}{16\pi^2}\left(-7\ln\frac{m_W}{\mu_t}+\frac{1}{3}+\frac{16}{9}\ln\frac{m_t}{\mu_t}+\frac{1}{3}\ln\frac{m_{H^+}}{\mu_t}\right).
\end{equation}

Next, we calculate the $\msbar$ masses of the gauge bosons as
\begin{equation}
 m_V^2(\mu_t)=m_V^{2,\os}+\bar\Sigma^T_V(m_V^2),
\end{equation}
with $V=W,Z$. Here, $\bar\Sigma^T_V(p^2)$ is the self energy for the
transverse mode with $1/\bar\varepsilon$ being subtracted. Here,
\begin{equation}
 \frac{1}{\bar\varepsilon}=\frac{2}{4-D}-\gamma_E+\ln4\pi,
\end{equation}
where $D$ is the spacetime dimension and $\gamma_E$ is the Euler
number. The superscript, $\os$, indicates the on-shell mass.

Using these, the Weinberg angle is calculated as
\begin{align}
 \cos\theta_W(\mu_t)&=\frac{m_W(\mu_t)}{m_Z(\mu_t)}.
\end{align}

Then, the $\msbar$ gauge couplings are given by
\begin{align}
 g_Y(\mu_t)&=\frac{e(\mu_t)}{\cos\theta_W(\mu_t)},\\
 g_2(\mu_t)&=\frac{e(\mu_t)}{\sin\theta_W(\mu_t)}.
\end{align}

Finally, the strong coupling constant is evaluated with {\tt RunDec}
\cite{Chetyrkin:2000yt,Herren:2017osy}. Notice that there are no one-loop
threshold corrections from the additional Higgs bosons.

For the later convenience, let us define
\begin{equation}
 v(\mu_t)=\frac{2\sin\theta_W(\mu_t)}{e(\mu_t)}m_W(\mu_t).
\end{equation}
\subsection{Yukawa Couplings}
The $\msbar$ tau mass is obtained from
\begin{equation}
 m_\tau(\mu_t)=m_\tau^\os\left[1+\bar\Sigma^S_\tau(m^2_\tau)+\frac{1}{2}\bar\Sigma^L_\tau(m^2_\tau)+\frac{1}{2}\bar\Sigma^R_\tau(m^2_\tau)\right],
\end{equation}
where $\bar\Sigma^S_\tau(p^2)$, $\bar\Sigma^L_\tau(p^2)$ and
$\bar\Sigma^R_\tau(p^2)$ are the scalar, the left-handed and the
right-handed parts of the self energy with $1/\bar\varepsilon$ being
subtracted.

As for the $\msbar$ masses of the top quark and the bottom quark, we
include the four-loop QCD corrections by using {\tt RunDec}
\cite{Chetyrkin:2000yt,Herren:2017osy}.  Then, we add the non-QCD one-loop
threshold corrections to the output of {\tt RunDec} as
\begin{equation}
 m_f(\mu_t)=m_f^{\tt RunDec}(m_t)\left[1+\bar\Sigma^S_{f,g_3=0}(m_f^2)+\frac{1}{2}\bar\Sigma^L_{f,g_3=0}(m_f^2)+\frac{1}{2}\bar\Sigma^R_{f,g_3=0}(m_f^2)\right],
\end{equation}
for $f=t,b$. Here, $m_f^{\tt RunDec}(m_t)$ is the output of {\tt RunDec}
and the subscript $g_3=0$ indicates that the strong coupling is switched
off in the calculation.

Then, the $\msbar$ Yukawa couplings are given by
\begin{align}
 y_t(\mu_t)&=\frac{\sqrt{2}}{\sin\beta(\mu_t)}\frac{m_t(\mu_t)}{v(\mu_t)},\\
 y_b(\mu_t)&=\frac{\sqrt{2}}{\cos\beta(\mu_t)}\frac{m_b(\mu_t)}{v(\mu_t)},\\
 y_\tau(\mu_t)&=\frac{\sqrt{2}}{\cos\beta(\mu_t)}\frac{m_\tau(\mu_t)}{v(\mu_t)}.
\end{align}
\subsection{Higgs Quartic Couplings}
To adjust the Higgs VEVs order by order in perturbative expansions, we
extend the scalar potential with tadpole terms as
\begin{equation}
 V_{\rm THDM}\to V_{\rm THDM}+T_hh+T_HH,
\end{equation}
where $T_h$ and $T_H$ are zero at the tree level.

The $\msbar$ values of these couplings are chosen as
\begin{equation}
 T_X(\mu_t)=\bar\Gamma_X^{(\rm tad)},
\end{equation}
with $X=h,H$. Here, $\bar\Gamma_X^{(\rm tad)}$ is the tadpole
contributions to the effective action with $1/\bar\varepsilon$ being
subtracted.

As for the scalars, the $\msbar$ masses are given by
\begin{equation}
 m^2_X(\mu_t)=m_X^{2,\os}+\bar\Sigma(m_X^2),
\end{equation}
with $X=h,H,A,H^+$.  Here, $\bar\Sigma_X(p^2)$ is the self energy with
$1/\bar\varepsilon$ being subtracted.

The $\msbar$ Higgs quartic couplings are then obtained as
\begin{align}
 \lambda_1&=\frac{1}{2v^2\cos^2\beta}\left[m_h^2+m_H^2-(1-\cos2\beta)m_A^2+(m_H^2-m_h^2)\cos2\alpha\right]\nonumber\\
&\hspace{3ex}+\frac{1}{2v^3\cos\beta}\left[(3\cos\alpha-\cos(\alpha-2\beta))T_H-(3\sin\alpha-\sin(\alpha-2\beta))T_h\right],\\
\lambda_2&=\frac{1}{2v^2\sin^2\beta}\left[m_h^2+m_H^2-(1+\cos2\beta)m_A^2-(m_H^2-m_h^2)\cos2\alpha\right]\nonumber\\
&\hspace{3ex}+\frac{1}{2v^3\sin\beta}\left[(3\sin\alpha+\sin(\alpha-2\beta))T_H+(3\cos\alpha+\cos(\alpha-2\beta))T_h\right],\\
\lambda_3&=\frac{1}{v^2}\left[2m_{H^+}^2-m_A^2+(m_H^2-m_h^2)\frac{\sin2\alpha}{\sin2\beta}\right]\nonumber\\
&\hspace{3ex}+\frac{1}{2v^3\sin2\beta}\nonumber\\
 &\hspace{5ex}\times\left[(3\sin(\alpha+\beta)+\sin(\alpha-3\beta))T_H+(3\cos(\alpha+\beta)+\cos(\alpha-3\beta))T_h\right],\\
 \lambda_4&=\frac{2}{v^2}(m_A^2-m_{H^+}^2),
\end{align}
where all the quantities appearing in the right-hand side are the
$\msbar$ values; we suppressed the renormalization scale, $\mu_t$, for
visibility.

\section{Generation of Data Points}\label{apx_mc}
In our analysis, we need to generate data points that consist of
$(m_H,m_A,\tan\beta,\cos(\beta-\alpha))$ for a fixed $m_{H^+}$. We first
take a random $\tan\beta$, which is uniformly distributed in the ranges
defined in Eqs.~\eqref{eq_fl_begin}-\eqref{eq_fl_end}. The other
variables are generated with the procedure described in this
section. The generated data points are then filtered by the perturbative
unitarity conditions and are passed to the next analysis. In this
appendix, we answer the following questions: (i) what is the appropriate
range for $m_H$, $m_A$ and $\cos(\beta-\alpha)$ that covers all the
points allowed by the perturbative unitarity? (ii) how can we
effectively generate data points that are allowed by the perturbative
unitarity?

A naive answer to question (i) is that
$|m_{H^+}^2-m_{H,A}^2|\lesssim8\pi v^2$ and $|\cos(\beta-\alpha)|\leq1$,
where $v\simeq 246~{\rm GeV}$. However, they are too weak to be used for
the parameter scan.  As we can see from Fig.~\ref{fig_scat1}, the
allowed mass differences are smaller than about $200~{\rm
GeV}$. However, one realizes that $\sqrt{8\pi}v\simeq1.2~\rm{TeV}$. It
also means that $H$ can be as light as $h$ and the mixing angle can
become large, which is why we naively expect no constraint on the mixing
angle. However, the allowed $|\cos(\beta-\alpha)|$ is smaller than about
$0.02$ and becomes much smaller in the large $\tan\beta$ regime. Thus,
if we scattered the data points over this naive range, we could get only
a very few points that satisfy the low energy constraints. That is why
we have question (ii).

\subsection{Necessary Conditions for Perturbative Unitarity}
Let us first analyze the perturbative unitarity conditions.
For arbitrary real numbers $A,B$ and $C$, the inequality,
\begin{equation}
 |A\pm\sqrt{B^2+C^2}|<1,
\end{equation}
can be reduced to
\begin{equation}
 |C|<\sqrt{(|A|-1)^2-B^2}~\&~|A|<1~\&~|B|<1-|A|.
\end{equation}
Applying it to the perturbative unitarity constraints, we get
constraints on $\lambda_1$ and $\lambda_2$ as
\begin{align}
 |\lambda_1|&<\mathcal T,\\
 |\lambda_2|&<\mathcal T,\\
 |\lambda_1+\lambda_2|&<\frac{2}{3}\mathcal T,\\
 |\lambda_1+\lambda_2|&<\mathcal T+\frac{\lambda_1\lambda_2}{\mathcal T},\\
 |\lambda_1+\lambda_2|&<\frac{\mathcal T}{3}+\frac{3\lambda_1\lambda_2}{\mathcal T},
\end{align}
which can be reduced to
\begin{align}
 |\lambda_1|&<\frac{\mathcal T}{3},\\
 |\lambda_2|&<\frac{\mathcal T}{3}.
\end{align}
Here, $\mathcal T=8\pi$.
As for $\lambda_3$ and $\lambda_4$, the constraints have the form of
\begin{align}
 a_i\lambda_3+b_i\lambda_4&<c_i,\label{eq_lp}
\end{align}
Here, $a_i$'s and $b_i$'s are constants and $c_i$'s are functions of
$\lambda_1$ and $\lambda_2$.

The range of $\lambda_3$ and $\lambda_4$ satisfying Eq.~\eqref{eq_lp}
can be determined by the simplex method of linear programming. We
consider simultaneous equations given by
\begin{align}
 a_i\lambda_3+b_i\lambda_4+z_i&=c_i,
\end{align}
where $z_i$'s are the slack variables. Then, we solve them under the
constraint of $z_k=z_l=0$ for each pair of $(k,l)$.  The solutions
satisfying $z_i\geq0$ correspond to the corners of the allowed region.
We search for such solutions and get
\begin{align}
 |\lambda_3|&<\frac{\mathcal T+\sqrt{\mathcal T^2-3\mathcal T|\lambda_1+\lambda_2|+9\lambda_1\lambda_2}}{3}\leq\frac{2}{3}\mathcal T,\\
 |\lambda_4|&<\frac{2}{3}\mathcal T,\\
 |\lambda_3+\lambda_4|&<\frac{\mathcal T+\sqrt{\mathcal T^2-3\mathcal T|\lambda_1+\lambda_2|+9\lambda_1\lambda_2}}{3}\leq\frac{2}{3}\mathcal T,\\
 |2\lambda_3+\lambda_4|&<\sqrt{\mathcal T^2-3\mathcal T|\lambda_1+\lambda_2|+9\lambda_1\lambda_2}\leq\mathcal T.
\end{align}

\subsection{Data Generation}
Let us go back to the problem of data generation.  
We define
\begin{align}
 \lambda_a&=\lambda_1\cos^2\beta-\lambda_2\sin^2\beta-(\lambda_3+\lambda_4)\cos2\beta,\\
 \lambda_b&=\lambda_1\cos^2\beta+\lambda_2\sin^2\beta+\frac{\lambda_4}{2},\\
 \lambda_c&=\lambda_1\cos^2\beta-\lambda_2\sin^2\beta+\lambda_3\cos2\beta.\label{eq_const_lam_c}
\end{align}
We will scatter $(\lambda_a,\lambda_b,\lambda_4)$ instead of
$(m_H,m_A,\cos(\beta-\alpha))$.  

We first generate random $(\lambda_a,\lambda_b)$. The scattering range
is given by
\begin{align}
 |\lambda_a|&<\frac{\mathcal T}{3}\left(1+2|\cos2\beta|\right),\\
 |\lambda_b|&<\frac{2}{3}\mathcal T,
\end{align}
which are derived from the inequalities in the previous subsection.  We
can further constrain the range with
\begin{align}
 m_{H^+}^2-m_h^2+\lambda_bv^2&=m_H^2>0.\label{eq_rel_mH}
\end{align}

Then, we calculate
\begin{align}
  \sin2(\beta-\alpha)&=\frac{\lambda_av^2\sin2\beta}{m_{H^+}^2-2m_h^2+\lambda_bv^2},\\
  \lambda_c&=\frac{2\cos2\beta}{v^2}\left[m_{H^+}^2 + \cos2(\beta-\alpha) \left(m_{H^+}^2 - 2 m_h^2 + \lambda_b v^2\right)\right]\nonumber\\
 &\hspace{3ex}-\lambda_a \cos4\beta,
\end{align}
with the assumption of $\cos2(\beta-\alpha)<0$\footnote{We could not
find any allowed points for the opposite case.}, and check
\begin{align}
 |\sin2(\beta-\alpha)|&<1,\\
 |\lambda_c|&<\frac{\mathcal T}{3}\left(1+2|\cos2\beta|\right),\\
 |\lambda_a-\lambda_c|&<\mathcal T|\cos2\beta|.
\end{align}
If any of them are not satisfied, we step back and regenerate
$(\lambda_a,\lambda_b)$.

Next, we generate a random $\lambda_4$. The scattering range is given by
\begin{align}
 |\lambda_4|&<\frac{2}{3}\mathcal T,\\
 m_{H^+}^2+\frac{\lambda_4}{2}v^2&=m_A^2>0,\label{eq_rel_mA}\\
\left|\frac{\lambda_a+2\lambda_b+\lambda_c}{2\sin^2\beta}-\lambda_4\right|&=|2\lambda_1\cot^2\beta|<\frac{2}{3}\mathcal T\cot^2\beta.
\end{align}
If all values of $\lambda_4$ have already been excluded, we go back
and regenerate $(\lambda_a,\lambda_b)$.

Finally, we calculate $m_H$ and $m_A$ using Eqs.~\eqref{eq_rel_mH} and
\eqref{eq_rel_mA}, and $\cos(\beta-\alpha)$ from
$\sin2(\beta-\alpha)$. Notice that we have assumed
$\cos2(\beta-\alpha)<0$ and $\sin(\beta-\alpha)>0$ in this
analysis. Then, we output $(m_H,m_A,\tan\beta,\cos(\beta-\alpha))$.

We find that the speed of the data generation is fast enough and
$50\%-60\%$ of the generated data points satisfy the perturbative
unitarity conditions.

\section{Couplings and Partial Decay Widths of Heavy Higgs Bosons}\label{apx_hb}
In this appendix, we summarize couplings and the partial decay widths of
the Higgs bosons, which are used for the inputs of {\tt HiggsBounds}
\cite{Bechtle:2013wla,Bechtle:2013gu,Bechtle:2011sb,Bechtle:2008jh,Bechtle:2015pma}.
We use the results of
\cite{Djouadi:1995gv,Djouadi:2005gj,Spira:2016ztx}.  The tree level
couplings of the neutral Higgs bosons are shown in Table
\ref{tbl_coup}. They are normalized by the corresponding couplings of
the SM Higgs boson having the same mass as that of the decaying
particle.

\begin{table}[t]
 \begin{tabular}{c|cccccc}
\hline\hline
  \multicolumn{7}{c}{Neutral Higgs Couplings}\\
\hline\hline
  &$g_{XUU}$&$g_{XDD}$&$g_{XLL}$&$g_{XVV}$&$g_{XAZ}$&$g_{XH^{\pm}W^{\mp}}$\\
\hline
  $h$&$\frac{\cos\alpha}{\sin\beta}$&$-\frac{\sin\alpha}{\cos\beta}$&$-\frac{\sin\alpha}{\cos\beta}$&$\sin(\beta-\alpha)$&$\cos(\beta-\alpha)$&$\mp\cos(\beta-\alpha)$\\
  $H$&$\frac{\sin\alpha}{\sin\beta}$&$\frac{\cos\alpha}{\cos\beta}$&$\frac{\cos\alpha}{\cos\beta}$&$\cos(\beta-\alpha)$&$-\sin(\beta-\alpha)$&$\pm\sin(\beta-\alpha)$\\
  $A$&$\cot\beta$&$\tan\beta$&$\tan\beta$&$0$&$0$&$1$\\
\hline\hline\hline
 \end{tabular}
\caption{The SM-value normalized couplings of the neutral Higgs bosons
at the tree level. The up-type quarks, the down-type quarks, the
leptons, the vector bosons and the neutral Higgs bosons are represented
by $U$, $D$, $L$, $V$ and $X$, respectively. The couplings of $A$ to
fermions are the pseudo-scalar type and the others are the scalar type.}
\label{tbl_coup}
\end{table}

In the following, we use the running mass for the quark mass;
\begin{equation}
 m_q\equiv \left(\frac{m_X}{\mu_0}\right)^{-\frac{2\alpha_s(\mu_0)}{\pi}}m_q(\mu_0).
\end{equation}
where $X$ represents the decaying particle. The reference value
$m_q(\mu_0)$ is calculated with {\tt RunDec}
\cite{Chetyrkin:2000yt,Herren:2017osy} with $\mu_0=500~{\rm GeV}$.  We
define the following variables;
\begin{align}
 x_i^X&=\frac{4m_i^2}{m_X^2},~y_i=\frac{4m_i^2}{m_Z^2},\\
 \lambda(m_i^2,m_j^2;m_k^2)&=\left(1-\frac{m_i^2}{m_k^2}-\frac{m_j^2}{m_k^2}\right)^2-\frac{4m_i^2m_j^2}{m_k^4}.
\end{align}

The loop induced couplings are given by
\begin{align}
 g_{Xgg}&=\left|\frac{\sum_{f=t,b}g_{Xff}A_{1/2}^X(x_f^X)}{A_{1/2}^h(x_t^X)}\right|,\\
 g_{X\gamma\gamma}&=\left|\frac{\sum_{f=t,b,\tau}N_c^fQ_f^2g_{Xff}A_{1/2}^X(x_f^X)+g_{XWW}A_1^X(x_W^X)+g_{XH^+H^-}A^X_0(x_{H^+}^X)}{\frac{4}{3}A_{1/2}^h(x_t^X)+A_1^h(x_W^X)}\right|,\\
 g_{XZ\gamma}&=\left|\frac{\sum_{f=t,b,\tau}g_{Xff}\bar A_f^X(x_f^X,y_f)+g_{XWW}\bar A_W^X(x_W^X,y_W)+g_{XH^+H^-}\bar A_{H^\pm}(x_{H^+}^X,y_{H^+})}{\bar A_t^h(x_t^X,y_t)+\bar A_W^h(x_W^X,y_W)}\right|,
\end{align}
where
\begin{align}
 A^{h,H}_0&=-x[1-xf(x)],\\
 A^{h,H}_{1/2}&=2x[1+(1-x)f(x)],\\
 A^{h,H}_1&=-[2+3x+3x(2-x)f(x)],\\
 A^{A}_{1/2}&=2xf(x),
\end{align}
and
\begin{align}
 \bar A^{h,H}_{H^+}(x,y)&=\frac{\cos2\theta_W}{\cos\theta_W}I_1(x,y),\\
 \bar A^{h,H}_{f}(x,y)&=2N_c^f\frac{Q_f(I_3^f-2Q_f\sin^2\theta_W)}{\cos\theta_W}[I_1(x,y)-I_2(x,y)],\\
 \bar A^{h,H}_{W}(x,y)&=\cos\theta_W\left\{4(3-\tan^2\theta_W)I_2(x,y)\right.\nonumber\\
 &\hspace{12ex}\left.+\left[\left(1+\frac{2}{x}\right)\tan^2\theta_W-\left(5+\frac{2}{x}\right)I_1(x,y)\right]\right\},\\
 \bar A^{A}_{f}(x,y)&=2N_c^f\frac{Q_f(I_3^f-2Q_f\sin^2\theta_W)}{\cos\theta_W}I_2(x,y).
\end{align}
Here, the tri-linear Higgs couplings are given by
\begin{align}
 g_{hH^+H^-}&=\frac{(m_h^2-2m_{H^+}^2)\cos(\alpha-3\beta)+(2m_{H^+}^2+3m_h^2-4m_A^2)\cos(\alpha+\beta)}{4m_{H^+}^2\sin2\beta},\\
 g_{HH^+H^-}&=\frac{(m_H^2-2m_{H^+}^2)\sin(\alpha-3\beta)+(2m_{H^+}^2+3m_H^2-4m_A^2)\sin(\alpha+\beta)}{4m_{H^+}^2\sin2\beta}.
\end{align}
Notice that $g_{AH^+H^-}=0$. The functions used in the above equations
are defined as
\begin{align}
 I_1(x,y)&=\frac{xy}{2(x-y)}+\frac{x^2y^2}{2(x-y)^2}[f(x)-f(y)]+\frac{x^2y}{(x-y)^2}[g(x)-g(y)],\\
 I_2(x,y)&=-\frac{xy}{2(x-y)}[f(x)-f(y)],
\end{align}
and
\begin{align}
 f(x)&=
\begin{cases}
 \arcsin^2\frac{1}{\sqrt{x}}&x\geq1\\
 -\frac{1}{4}\left[\log\frac{1+\sqrt{1-x}}{1-\sqrt{1-x}}-i\pi\right]^2&x<1
\end{cases},\\
 g(x)&=
\begin{cases}
 \sqrt{x-1}\arcsin\frac{1}{\sqrt{x}}&x\geq1\\
 \frac{\sqrt{1-x}}{2}\left[\log\frac{1+\sqrt{1-x}}{1-\sqrt{1-x}}-i\pi\right]&x<1
\end{cases}.
\end{align}

We include the following non-SM partial decay widths of the neutral
Higgs bosons;
\begin{align}
 \Gamma(H\to XX)&=\frac{G_Fm_Z^4}{16\sqrt{2}\pi m_H}g_{HXX}^2\sqrt{1-4\frac{m_X^2}{m_H^2}},\\
 \Gamma(X_i\to X_jZ)&=\frac{G_Fm_{X_i}^3}{8\sqrt{2}\pi}g_{X_iX_jZ}^2\lambda^{3/2}(m_Z^2,m_{X_j}^2;m_{X_i}^2),\\
 \Gamma(X\to H^- W^+)&=\Gamma(X\to H^+ W^-)\nonumber\\
 &=\frac{G_Fm_X^3}{8\sqrt{2}\pi}g_{XH^\pm W}^2\lambda^{3/2}(m_W^2,m_{H^+}^2;m_{X}^2),
\end{align}
where the relevant couplings are given by
\begin{align}
 g_{Hhh}&=\frac{\cos(\beta-\alpha)}{m_Z^2\sin2\beta}\left[(m_H^2+2m_h^2-3m_A^2)\sin2\alpha+m_A^2\sin2\beta)\right],\\
 g_{HAA}&=\frac{1}{2m_Z^2\sin2\beta}\left[(m_H^2-2m_A^2)\sin(\alpha-3\beta)+(3m_H^2-2m_A^2)\sin(\alpha+\beta))\right].
\end{align}

As for the charged Higgs boson, we consider the following partial decay
widths;
\begin{align}
 \Gamma(H^+\to XW^+)&=\frac{G_Fm_{H^+}^3}{8\sqrt{2}\pi}g_{XH^-W^+}^2\lambda^{3/2}(m_W^2,m_{X}^2;m_{H^+}^2),\\
 \Gamma(H^+\to tb)&=\frac{3G_Fm_{H^+}}{4\sqrt{2}\pi}\left[m_t^2\cot^2\beta+m_b^2\tan^2\beta\right]\lambda^{1/2}(m_t^2,m_b^2;m_{H^+}^2),\\
 \Gamma(H^+\to \tau\nu)&=\frac{G_Fm_{H^+}}{4\sqrt{2}\pi}m_\tau^2\tan^2\beta\left(1-\frac{m_\tau^2}{m_{H^+}^2}\right)^3.
\end{align}

\bibliographystyle{apsrev4-1}
\bibliography{thdm}
\end{document}